\documentclass[twocolumn,showpacs,preprintnumbers,amsmath,amssymb]{revtex4}
\usepackage{graphicx}
\usepackage{dcolumn}
\usepackage{bm}

\newcommand{\bef}{\begin{figure}}
\newcommand{\eef}{\end{figure}}

\newcommand{\be}{\begin{equation}}
\newcommand{\ee}{\end{equation}}
\newcommand{\bea}{\begin{eqnarray}}
\newcommand{\eea}{\end{eqnarray}}
\newcommand {\tria}{\varepsilon_{3}}
\newcommand {\ecc}{\varepsilon_{2}}

\DeclareMathOperator{\atantwo}{atan2} 

\newcommand {\mean}[1]{\left\langle #1 \right\rangle}
\begin{document}

\title{Multiplicity, average transverse momentum and azimuthal anisotropy in U+U collisions at $\sqrt{s_{\mathrm NN}}$ = 200 GeV using AMPT model}

\author{Md. Rihan Haque$^1$, Zi-Wei Lin$^2$, and Bedangadas Mohanty$^1$ }
\affiliation{$^1$Variable Energy Cyclotron Centre, Kolkata 700064, India and  $^2$Department of Physics, East Carolina University, Greenville, NC 27858-4353, USA}

\date{\today}
\begin{abstract}

Using a multi-phase transport (AMPT) model that includes the implementation of
deformed Uranium nuclei, we have studied the centrality dependence of the charged particle 
multiplicity ($N_{\rm {ch}}$, $dN_{\rm {ch}}/d\eta$),  average transverse momentum ($\langle p_{\rm T} \rangle$), 
eccentricity ($\ecc$), triangularity ($\tria$), their fluctuations, elliptic flow ($v_{\rm 2}$) and 
triangular flow ($v_{\rm 3}$) for different configurations 
of U+U collisions at midrapidity for $\sqrt{s_{\rm {NN}}}$ = 200 GeV. The calculations have been done for both 
the default and string melting versions of the AMPT model. The results are compared to the corresponding observations 
from Au+Au collisions. We find that for the U+U collisions the $dN_{\rm {ch}}/d\eta$ at midrapidity is enhanced by 
about 15-40\% depending on the collision and model configuration chosen, compared to Au+Au collisions. 
Within the several configurations studied, the tip-to-tip collisions leads to the largest values of $N_{\rm {ch}}$,
transverse energy ($E_{\rm {T}}$) and $\langle p_{\rm T} \rangle$. The $\langle \ecc \rangle$  and its fluctuation 
shows a rich centrality dependence, whereas not much variations are observed for $\langle \tria \rangle$  and its 
fluctuations. The U+U side-on-side collision configuration provides maximum values of $\langle \ecc \rangle$ 
and minimum values of eccentricity fluctuations, whereas for peripheral collisions and mid-central collisions 
minimum values of $\langle \ecc \rangle$ and maximum value of eccentricity fluctuations are observed for body-to-body configuration 
and the tip-to-tip configuration has minimum value of $\langle \ecc \rangle$  
and maximum value of eccentricity fluctuations for central collisions.  The calculated $v_{\rm 2}$ closely correlates with the eccentricity
in the model. It is smallest for the body-to-body configuration in peripheral and mid-central collisions while it is minimum for tip-to-tip 
configuration in central collisions. For peripheral collisions the $v_{\rm 2}$ in U+U  can be about 40\% larger than in Au+Au 
whereas for central collisions it can be a factor 2 higher depending on the collision configuration. It is also observed that 
the $v_{\rm 3}$($p_{\rm T}$) is higher for tip-to-tip and body-to-body configurations compared to other systems for the collision centrality studied. 

\end{abstract}
\maketitle

\section{INTRODUCTION}

In Au+Au collisions at the Relativistic Heavy Ion Collider facility, 
large values of elliptic flow and large suppression in high transverse momentum 
hadron production relative to the $p$+$p$ collisions have been reported~\cite{rhicwhitepapers}. 
The dominant interpretation of these measurements have indicated that the relevant degrees of freedom 
in these collisions are quarks and gluons. Deformed nuclei collisions such as U+U will allows us
to investigate the initial conditions, hydrodynamic behavior, path length dependence of partonic energy loss~\cite{heniz,hirano,bm}, 
possible local parity violation~\cite{voloshin} and other physics topics beyond what we have learned from Au+Au collisions.
The commissioning of the Electron Beam Ion Sources~\cite{ebis} will enable RHIC to collide Uranium ions. 
U+U collisions are being planned for 2012 with center of mass energy around 200 GeV~\cite{bnlpac}. 

In contrast to central Au+Au collisions, because of the prolate shape of Uranium, there
are configurations (e.g body-to-body, defined later) in which central U+U collisions are not spherical in the
transverse plane, but has an elliptic shape. At RHIC we have
observed an increase in $v_{\rm 2}$/$\ecc$ with increase in transverse particle
density~\cite{rhicwhitepapers}. This corresponds to the dilute regime predictions in kinetic theory~\cite{heiselberg}. 
For the hydrodynamic regime, one expects $v_{\rm 2}$/$\ecc$ to saturate
with increase in transverse particle density~\cite{heiselberg}. One way to extend the transverse particle density beyond
what has been achieved at RHIC is by performing U+U collisions or going to higher beam
energies as at LHC. Studies suggest that the maximum transverse particle density attended in U+U collisions 
could be about 6\%-35\% higher than Au+Au collisions depending on the colliding configuration~\cite{heniz,hirano,declan}. 
Furthermore, several possible configurations of U+U collisions can occur,
depending on the angles of the two incoming Uranium nuclei relative to the reaction plane. This will help
in constraining the initial condition models by the measurement of $v_{\rm 2}$, $v_{\rm 3}$
and their fluctuations in U+U and comparing the same to the corresponding results in Au+Au collisions.
Galuber-based model simulations suggest an increased value of $\langle \ecc \rangle$  (up to 30\%) and 
eccentricity fluctuations in deformed U+U collisions relative to Au+Au collisions~\cite{hiroshi,bm}.

Furthermore it has been shown from the space-time evolution of high energy non-central symmetric heavy ion collisions
using  relativistic hydrodynamics that the matter expands preferentially in the impact
parameter direction and the expanding shells leave a rarefaction behind. As a consequence of early pressure 
gradient this could at freeze-out lead to three distinct fireballs being produced. This was referred to as the 
nutcracker scenario~\cite{teaney}. Subsequently it has been pointed out that such a phenomena is missing 
for U+U collisions due to the time evolution of the initial transverse energy density profile within a hydrodynamical 
frame work~\cite{klob}.

The energy loss of partons in a hot and dense colored QCD matter depends not only on the medium density and
color factor but also on the path length traversed by the parton. Theories of energy loss for fast partons 
support a non-linear dependence of parton energy loss on the path-length, but this has not yet been 
fully tested in experiment, due to the small difference in path lengths for the parton traversing in-plane and 
out-of-plane for Au+Au collisions. Body-to-body U+U collisions are expected to provide almost twice 
as much difference between the in-plane and out-of-plane path lengths for the same eccentricity as
semi-peripheral Au+Au collisions. This in turn is expected to increase by 100\% the absolute
value of radiative energy loss and its difference between in-plane and out-of-plane
directions~\cite{heniz}.

Parity is conserved globally in the strong interaction, but local parity violation 
is possible because of the topological structure of QCD~\cite{kharzeev}. It has been proposed that
heavy-ion collisions at high energies, provide an unique opportunity to observe
local parity violation~\cite{finch}. The magnetic field required for the parity violating
signal exists in non-central heavy-ion collisions and is produced due to the
spectators. In central U+U body-to-body collisions, there are no spectators 
(small or zero magnetic field), while in certain configurations the geometry of the 
collision zone induces finite $v_{\rm 2}$. Background process to local parity violation 
are expected to be related to $v_{\rm 2}$, while the signal is expected to be related to 
the magnetic field strength~\cite{voloshin}. One can then use a comparative study of 
local parity violation observables for Au+Au and U+U collisions at 
similar energies to interpretate the measurements~\cite{starlpv}. 

Most of the previous model based study of U+U collisions have made use of Monte Carlo
Galuber simulation~\cite{bm, hiroshi} or it is coupled to a hydrodynamic evolution~\cite{heniz,hirano}. 
Some investigations exists for selecting special orientations of U+U collisions using event generators~\cite{declan}.
In this work, we mainly focus on centrality dependence of 
$dN_{\rm {ch}}/d\eta$, $\langle p_{\rm T} \rangle$, $\ecc$, $\tria$,
their fluctuations, $v_{\rm 2}$ and $v_{\rm 3}$ for several configurations of U+U collisions. 
The results are also compared to corresponding observations in Au+Au collisions.

The paper is organized as follows. In the next section we discuss the implementation of 
U+U collision in the AMPT model~\cite{ampt}. We also discuss the specific configurations of U+U 
collisions that we study in this paper. Section III presents the results, 
which includes the $N_{\rm {ch}}$, $dN_{\rm {ch}}/d\eta$, $E_{\rm T}$ and  $\langle p_{\rm T} \rangle$. 
This is followed by discussion on several geometrical variables like $\ecc$, eccentricity fluctuations, 
$\tria$ and its fluctuations. Finally we present results on centrality and transverse momentum dependence 
of $v_{\rm 2}$ and $v_{\rm 3}$. All results are obtained using both the default and string melting
versions of the AMPT model~\cite{ampt} and the U+U results are compared to corresponding results from Au+Au
collisions. Finally in section IV we present a summary of our findings.

\section{Implementing Uranium collisions in AMPT}

\bef
\begin{center}
\includegraphics[scale=0.4]{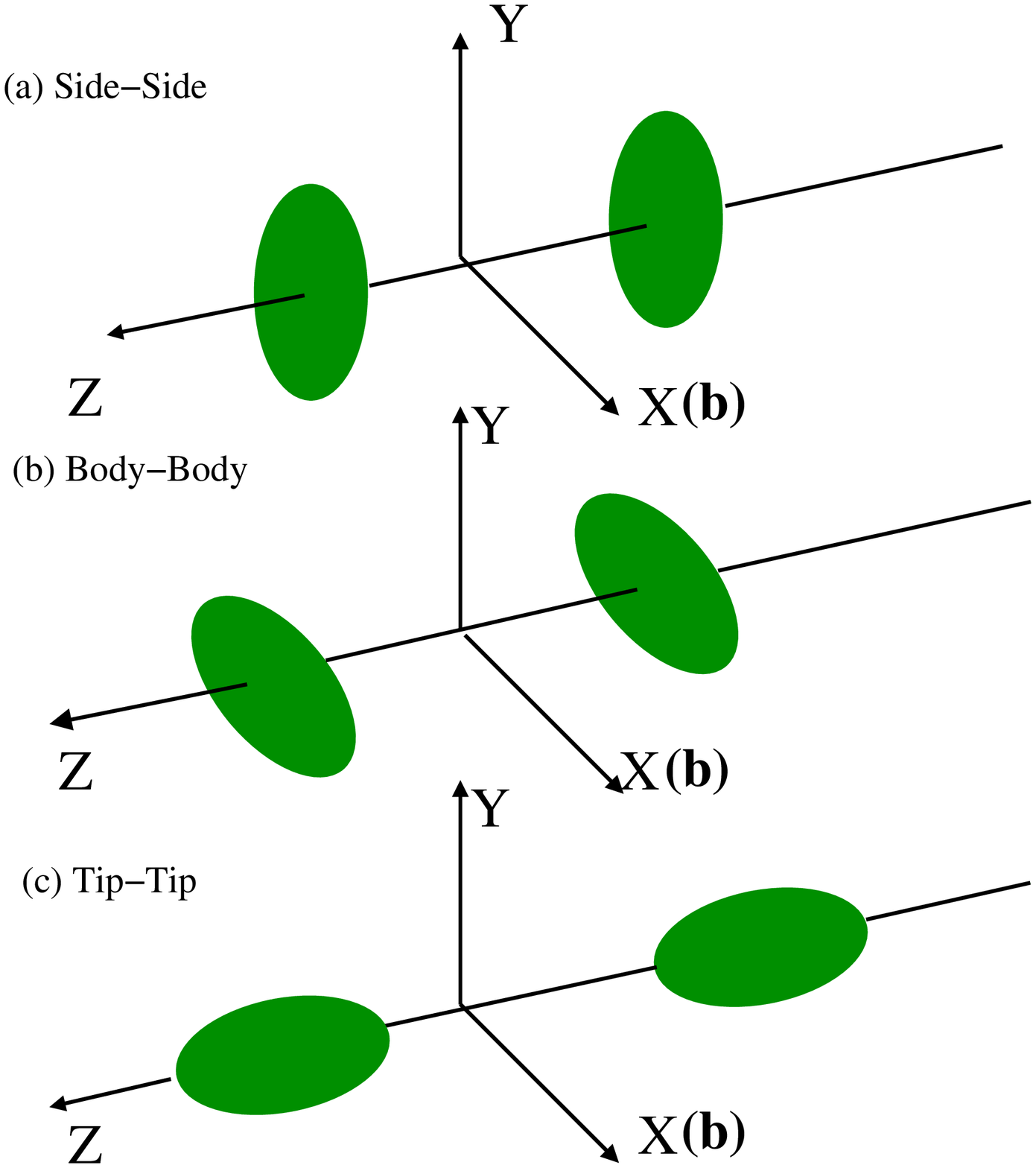}
\caption{(Color online) Different configurations of U+U collisions studied in the present work. The Z-axis
is the beam direction. X({\bf b}) represents that the impact parameter direction is along the X-axis. 
For more details refer to text and Table~\ref{table1}.}
\label{fig1}
\end{center}
\eef
In the current work, U+U collision is implemented in the AMPT model as follows.
The nucleon density distribution is parameterized as a deformed Woods-Saxon profile~\cite{Hagino:2006fj}
\begin{eqnarray}
\rho & = & \frac{\rho_0}{1 + \exp{([r - R']/a)}},  \label{eq:deformed_ws} \\
R' & = & R \left[1 + \beta_2Y_2^0(\theta) + \beta_4Y_4^0(\theta)\right],
\end{eqnarray}
where $\rho_0$ is the normal nuclear density, $R$ is the radius of the nucleus and $a$ denotes 
the surface diffuseness parameter.   We have used $R = 6.81$~fm and $a = 0.55$~fm for $^{238}$U nucleus. 
The $Y_l^m(\theta)$ denotes the spherical harmonics and $\theta$ is the polar angle 
 with the symmetry axis of the nucleus. 
Deformation parameters are $\beta_2 = 0.28$~\cite{heniz} and $\beta_4 = 0.093$~\cite{Moller:1993ed} for Uranium.
The presence of $\beta_4$ modifies the shape of Uranium compared to that only with $\beta_2$~\cite{heniz,declan}.
The radius increases $\sim$6\% (3\%) at $\theta = 0$ ($\theta = \pi/2$), while it decreases $\sim$3\% around $\theta = \pi/4$~\cite{bm}.
The positions of nucleons are sampled by $4\pi r^2 \sin{(\theta)} \rho(r) ~ d\theta d\phi$,
where the absolute normalization of $\rho(r)$ is irrelevant.
Both projectile and target U nuclei are randomly rotated along
the polar and azimuthal directions event-by-event with the probability 
 distribution $\sin{\Theta}$ and uniform distribution for $\Theta$ and $\Phi$, respectively.
The $\sin{\Theta}$ weight needs to be implemented to simulate unpolarized 
nucleus-nucleus collisions. 

In this work primarily three types of configuration of U+U collisions are studied and compared to
U+U collisions without any specific choice of orientation and Au+Au collisions. These specific 
configurations will be termed as body-to-body, side-on-side and tip-to-tip in the rest of the
paper and are shown in Fig.~\ref{fig1}. The details in terms of $\theta$ and $\phi$ angles of 
the orientation of the nuclei for these configurations studied in this paper are given in Table~\ref{table1}.

\begin{table}
\caption{The details of the angular configuration of U+U collisions used in this study. The
subscript $p$ and $t$ denotes the projectile and target respectively. In the simulations, for tip-to-tip
configuration the $\theta$ is varied as 0$\pm$0.07 radian; for body-to-body the $\theta$ is varied as
$\pi$/2$\pm$0.005 radian and $\phi$ as 0$\pm$0.0025 radian and for side-on-side, the $\theta$ 
and  $\phi$ are varied as $\pi$/2$\pm$0.005 and $\pi$/2$\pm$0.17 radians respectively.
\label{table1}}
\begin{tabular}{|l|c|c|c|c|c|r}
\tableline
Configuration& $\theta_p$ & $\theta_t$ & $\phi_p$ & $\phi_t$ & Impact parameter \\
\tableline
general        &  0--$\pi$              &  0--$\pi$          & 0--2$\pi$         & 0--2$\pi$    & random \\
tip-to-tip        &  0            &  0        & 0--2$\pi$         & 0--2$\pi$    & minor axis \\
body-to-body      &  $\pi$/2     &  $\pi$/2 & 0     & 0 & major axis \\
side-on-side      &  $\pi$/2     &  $\pi$/2 & $\pi$/2     & $\pi$/2 & minor axis \\
\tableline
\end{tabular}
\end{table}

The AMPT model takes initial conditions from HIJING~\cite{hijing}.
However the mini-jet partons are made to undergo scattering before they are allowed
to fragment or recombine into hadrons. The string melting version of the AMPT model 
(labeled here as SM) is based on the idea that for energy densities beyond 
a critical value of $\sim$ 1 GeV/$fm^{3}$, it is difficult to visualize the coexistence of 
strings (or hadrons) and partons. Hence the need to melt the strings to partons. This is done 
by converting the mesons to a quark and anti-quark pair, baryons to three quarks etc.  
The scattering of the quarks are based on parton cascade ZPC~\cite{ampt}. Once the interactions 
stop, the partons then hadronizes through the mechanism of parton coalescence. The interactions between
the mini-jet partons in default AMPT model and those between partons in the AMPT-SM
model could give rise to substantial $\langle v_{2} \rangle$. The parton-parton interaction 
cross section is taken as 10 mb. The results presented below uses both the default and SM
version of the AMPT model.

\section{RESULTS}

\subsection{Multiplicity, transverse energy and average transverse momentum}
\bef
\includegraphics[scale=0.35]{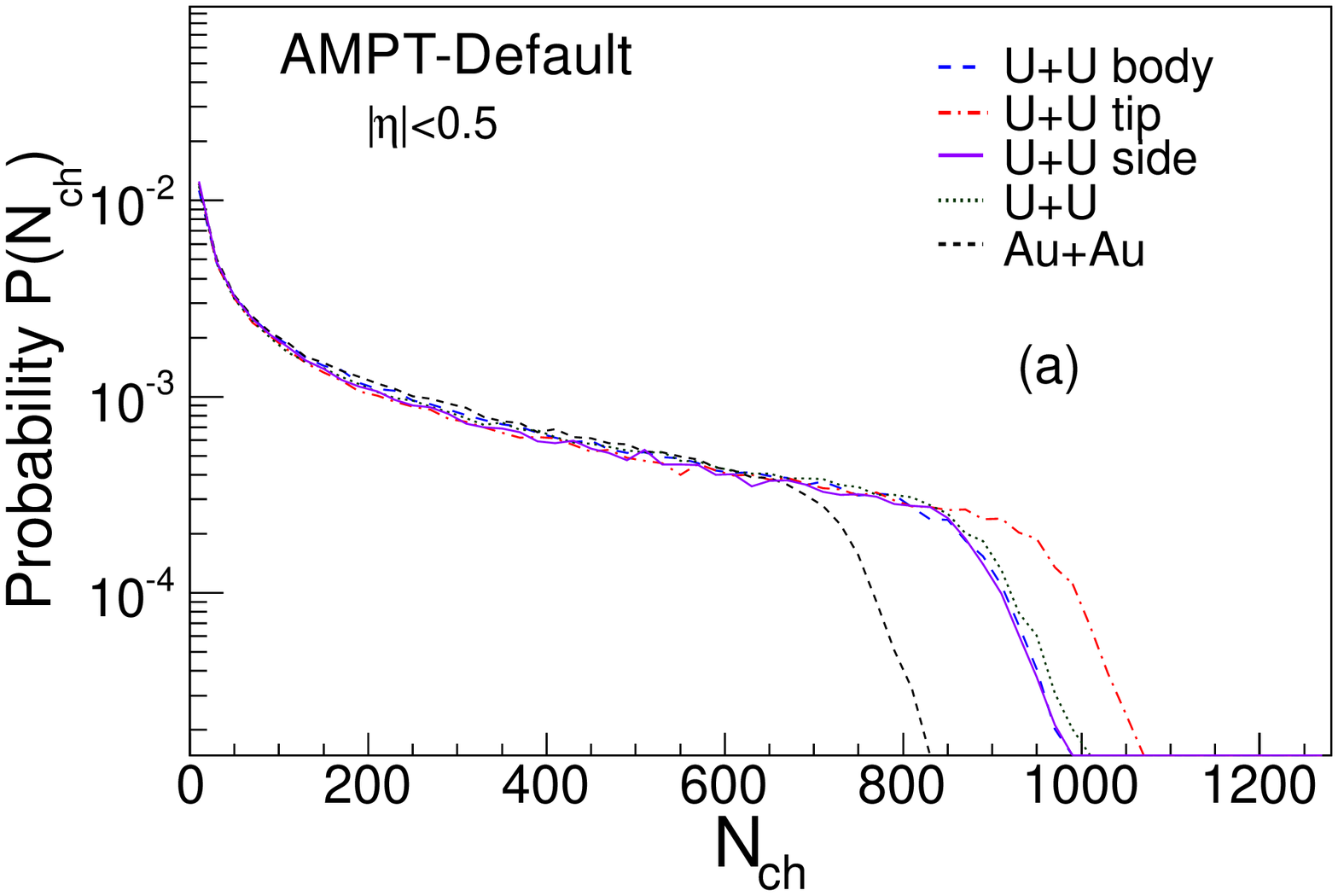}
\includegraphics[scale=0.35]{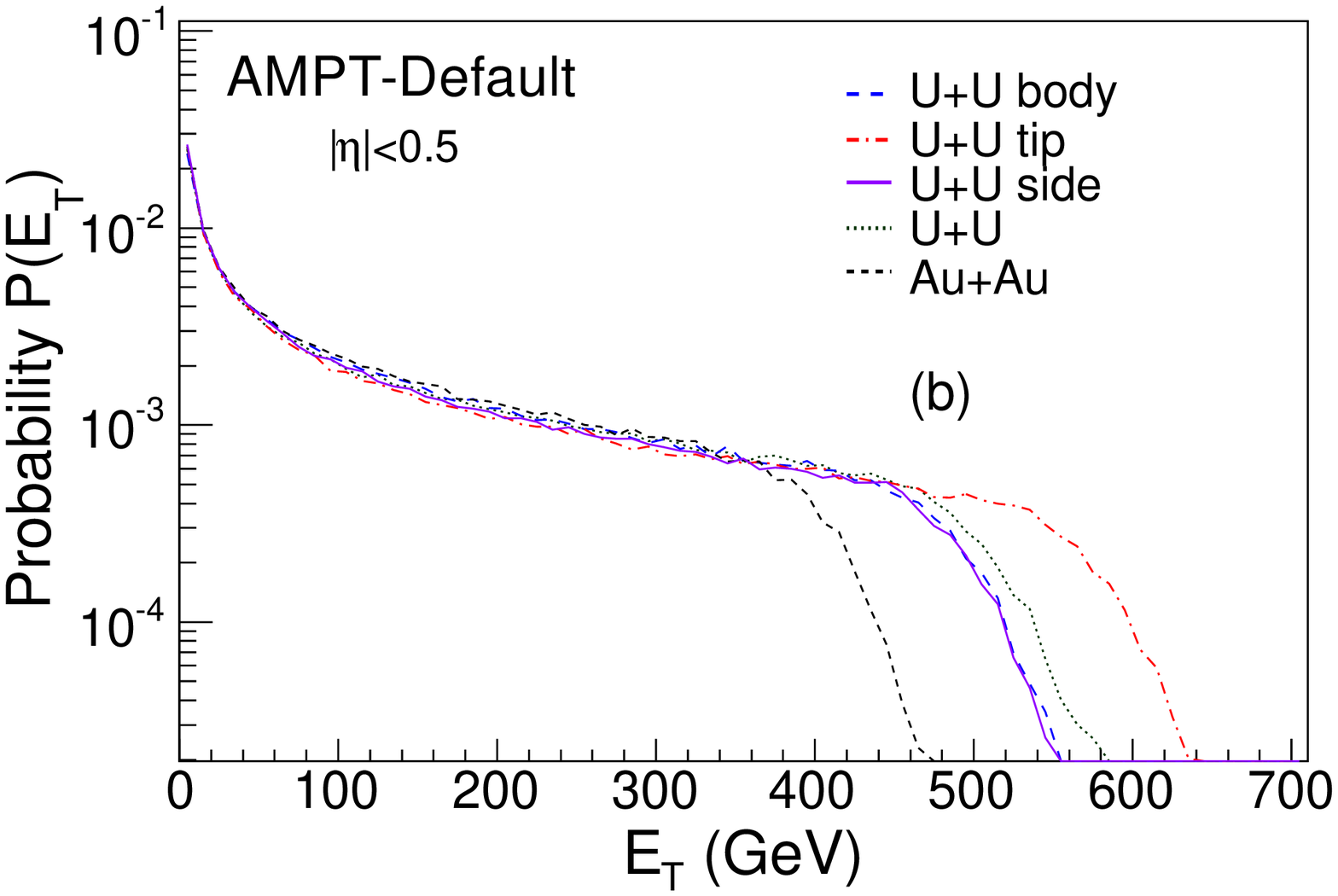}
\caption{(Color online) (a) Probability distribution of total charged particle
multiplicity ($N_{\rm {ch}}$) and (b) charged particle transverse energy 
($E_{\rm T}$). Both results are at midrapidity ($\mid \eta \mid$ $<$ 0.5) for minimum bias 
U+U collisions at $\sqrt{s_{\rm {NN}}}$ = 200 GeV from default AMPT model.
The different colored lines corresponds to different configurations of U+U
collisions. Also shown for comparison are the results from Au+Au collisions 
at $\sqrt{s_{\rm {NN}}}$ = 200 GeV from default AMPT model as short dashed lines.}
\label{fig3a}
\eef
\bef
\includegraphics[scale=0.35]{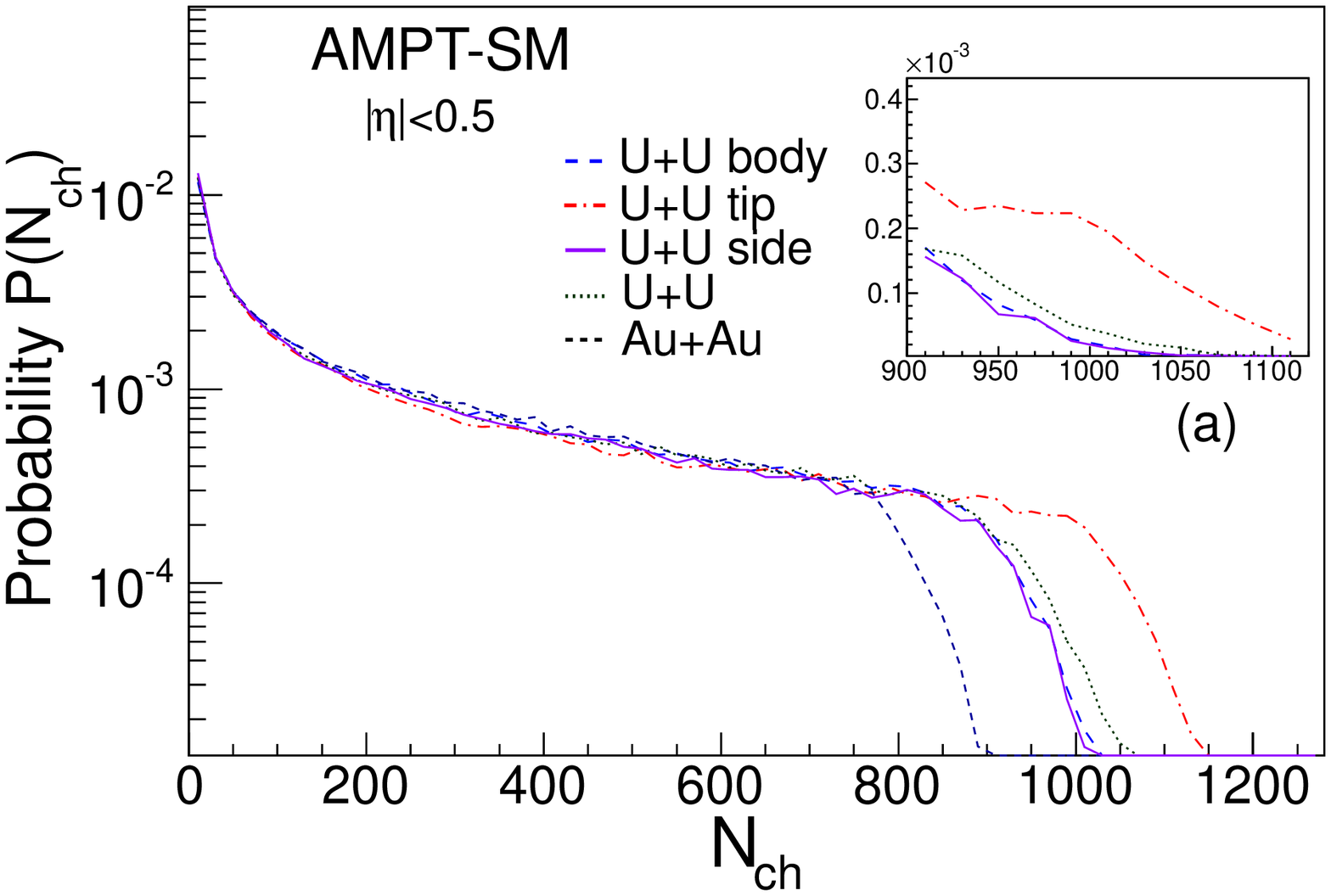}
\includegraphics[scale=0.35]{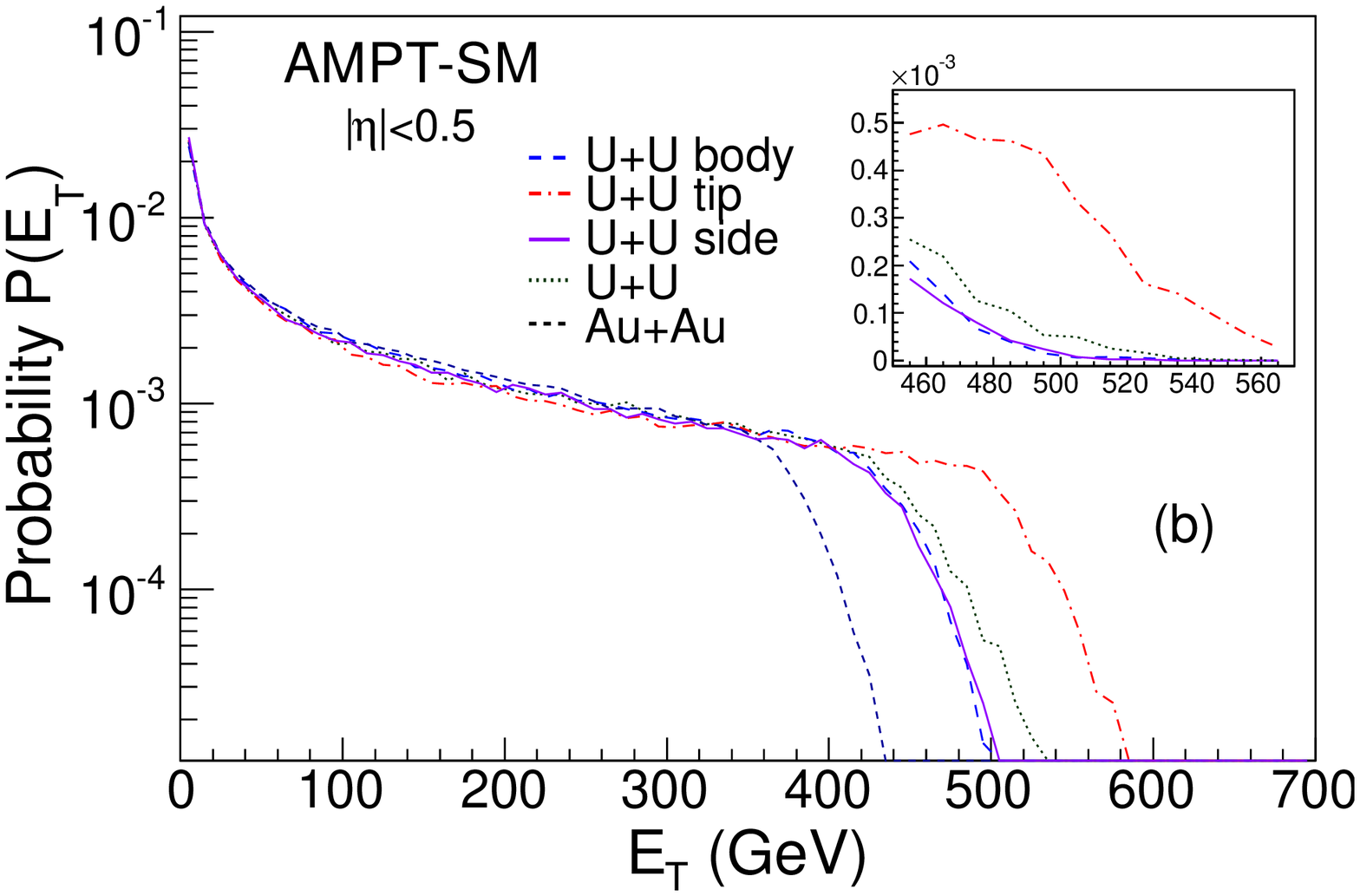}
\caption{(Color online) Same as Fig.~\ref{fig3a} for AMPT string
  melting version.  The inset shows the distributions for central
  collisions in an expanded scale.}
\label{fig3b}
\eef
Figure~\ref{fig3a} shows default AMPT model simulated minimum bias charged 
particle multiplicity ($N_{\rm {ch}}$) and 
charged particle transverse energy ($E_{\rm T}$) distributions 
at midrapidity ($\mid \eta \mid$ $<$ 0.5) in 
U+U collisions for different configurations for 
$\sqrt{s_{\rm {NN}}}$ = 200 GeV. For comparison also shown in
Fig.~\ref{fig3a} are the results from Au+Au (symmetric nuclei) 
collisions for the same kinematic conditions. The shapes of the distributions
are very similar for different configurations of U+U collisions and those
from Au+Au collisions. However the maximum values of $N_{\rm {ch}}$ 
and $E_{\rm T}$ attained for U+U collisions in various configurations 
is found to be about 15\%-35\% higher than the corresponding values 
from Au+Au collisions. The tip-to-tip configuration in U+U collisions
allows to attain the maximum $N_{\rm {ch}}$ and $E_{\rm T}$ values
among the various cases studied. The rest of the configurations for 
U+U seems to give similar values.

Figure~\ref{fig3b} shows the corresponding results using the string melting version of the 
AMPT model. The inset of the figure shows the distributions for
central collisions in an expanded scale.  In general the charged particle multiplicity is about 8\% higher compared to
the default case. However the transverse energy of charged particles at midrapidity is 
about 10\% lower for string melting case compared to default version of AMPT. The rest of
the trends for different configurations for U+U collisions relative to each other and
to the Au+Au collisions are similar in the both versions of the model.

\bef
\includegraphics[scale=0.35]{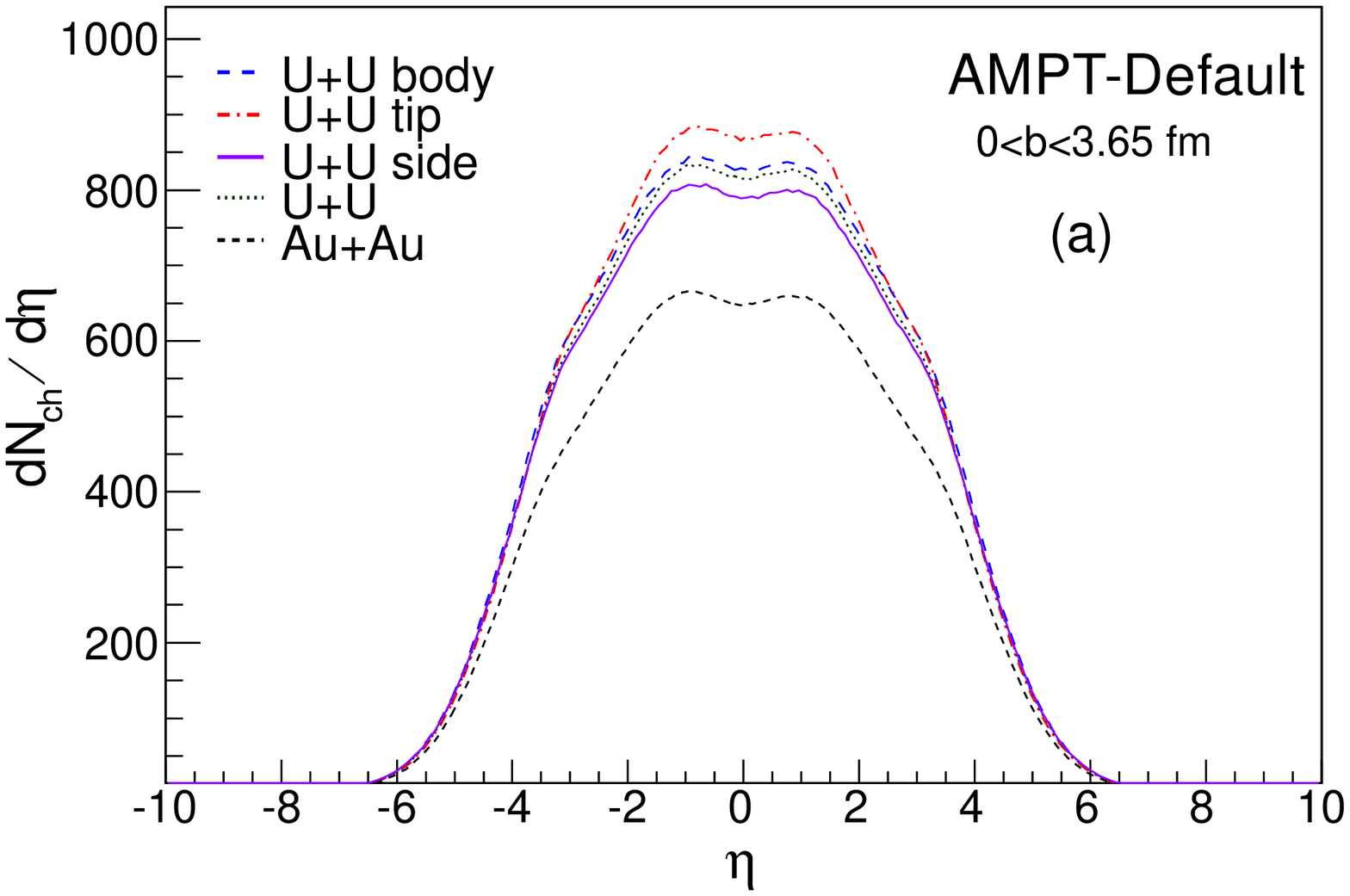}
\includegraphics[scale=0.35]{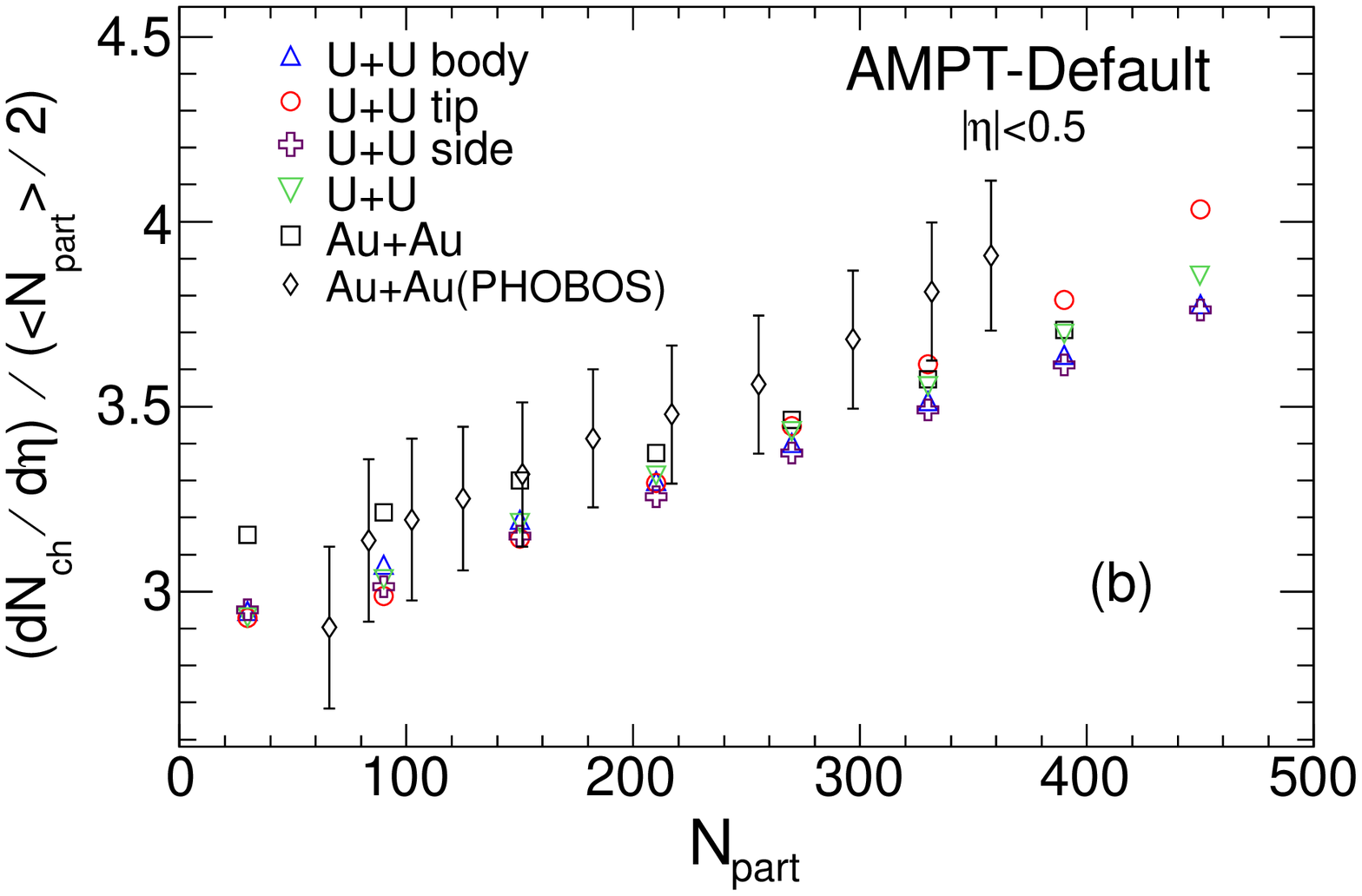}
\includegraphics[scale=0.35]{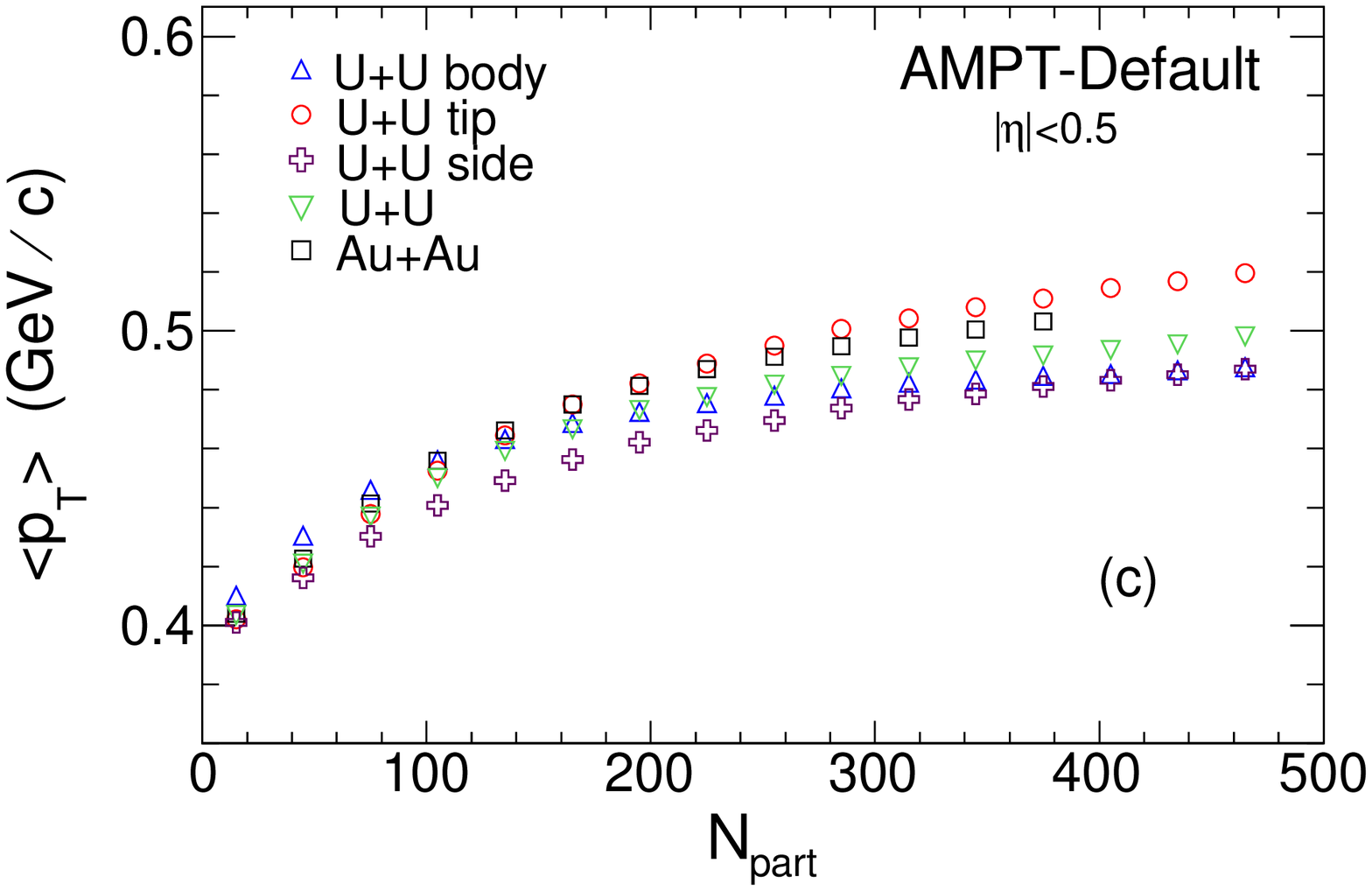}
\caption{(Color online) (a) Charged particle pseudorapidity ($dN_{\rm {ch}}/d{\eta}$) distribution for central collisions (impact parameter = 3.65 fm)
as a function of pseudorapidity ($\eta$) for U+U collisions with different
configurations of collision and Au+Au collisions at $\sqrt{s_{\rm {NN}}}$ = 200 GeV using default AMPT model. (b) $dN_{\rm {ch}}/d{\eta}$ per participating
nucleon ($N_{\rm {part}}$) pair versus  $N_{\rm {part}}$ at midrapidity 
($\mid \eta \mid < 0.5$) for above cases.
Also shown are the experimental results from Au+Au collisions from PHOBOS
experiment at RHIC~\cite{phobos}. 
(c) Average transverse momentum ($\langle p_{\rm T} \rangle$) 
of charged particles as a function of $N_{\rm {part}}$ at midrapidity 
($\mid \eta \mid < 0.5$) for above collision configurations.}
\label{fig4a}
\eef

\bef
\includegraphics[scale=0.35]{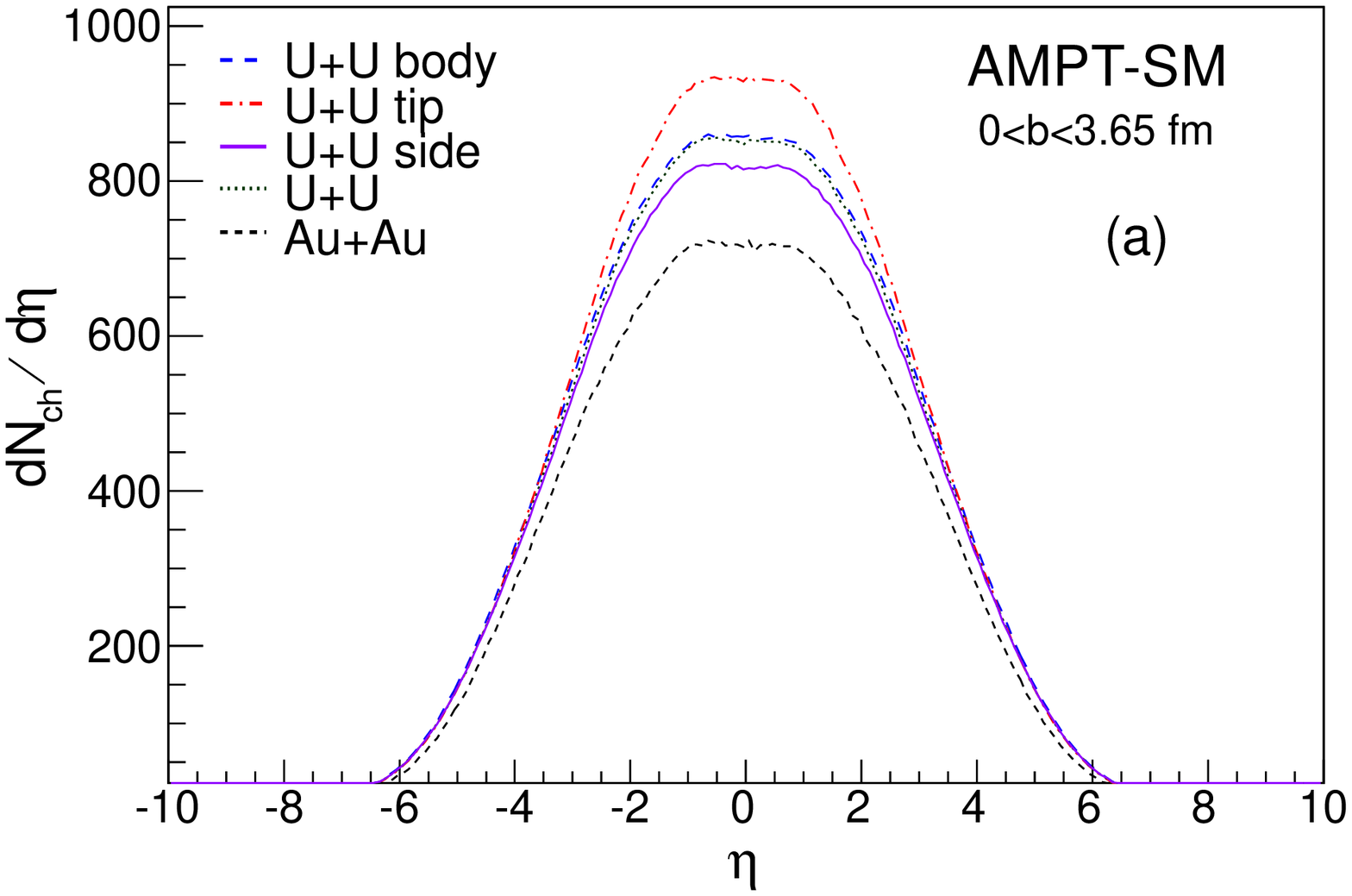}
\includegraphics[scale=0.35]{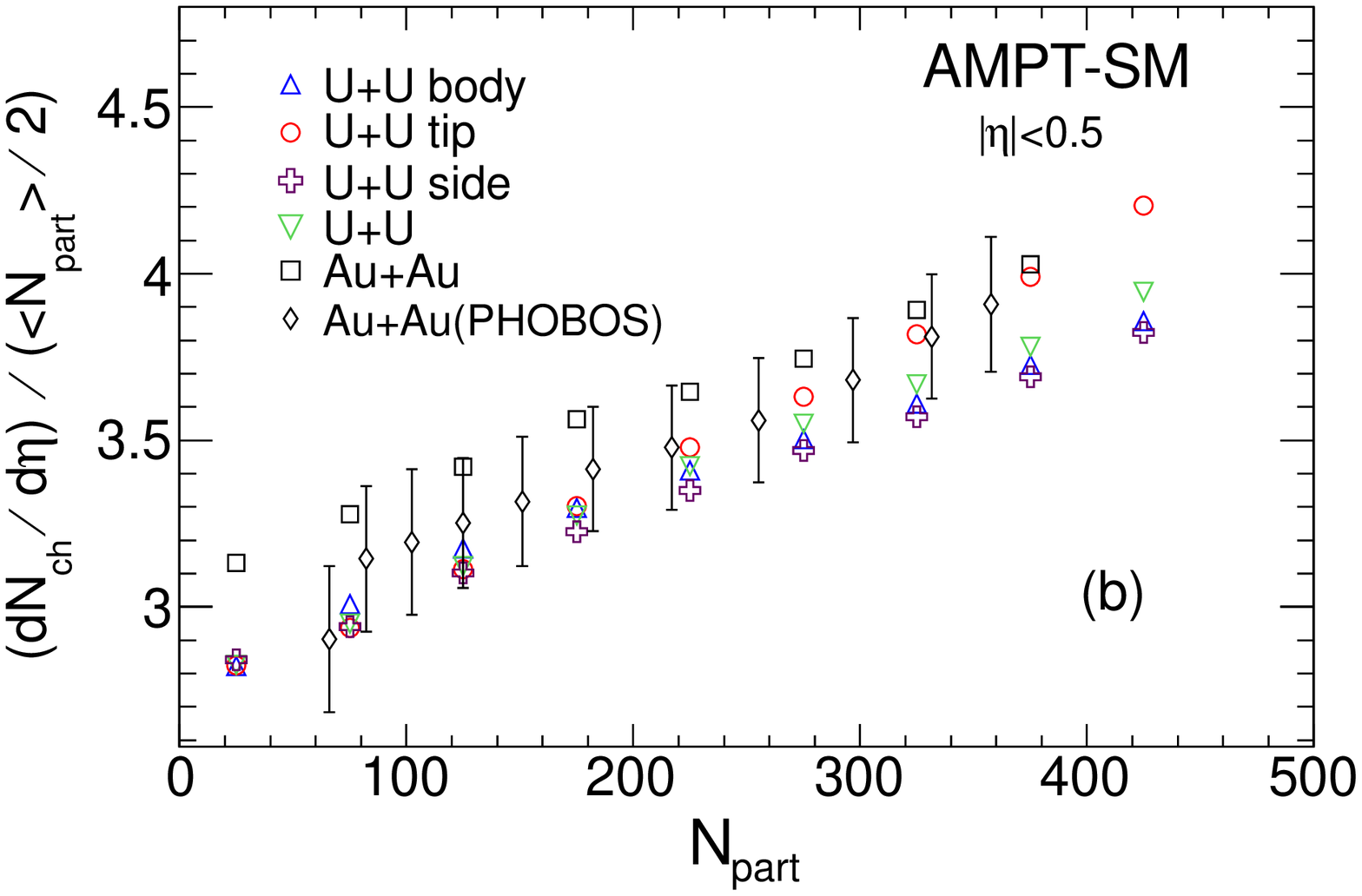}
\includegraphics[scale=0.35]{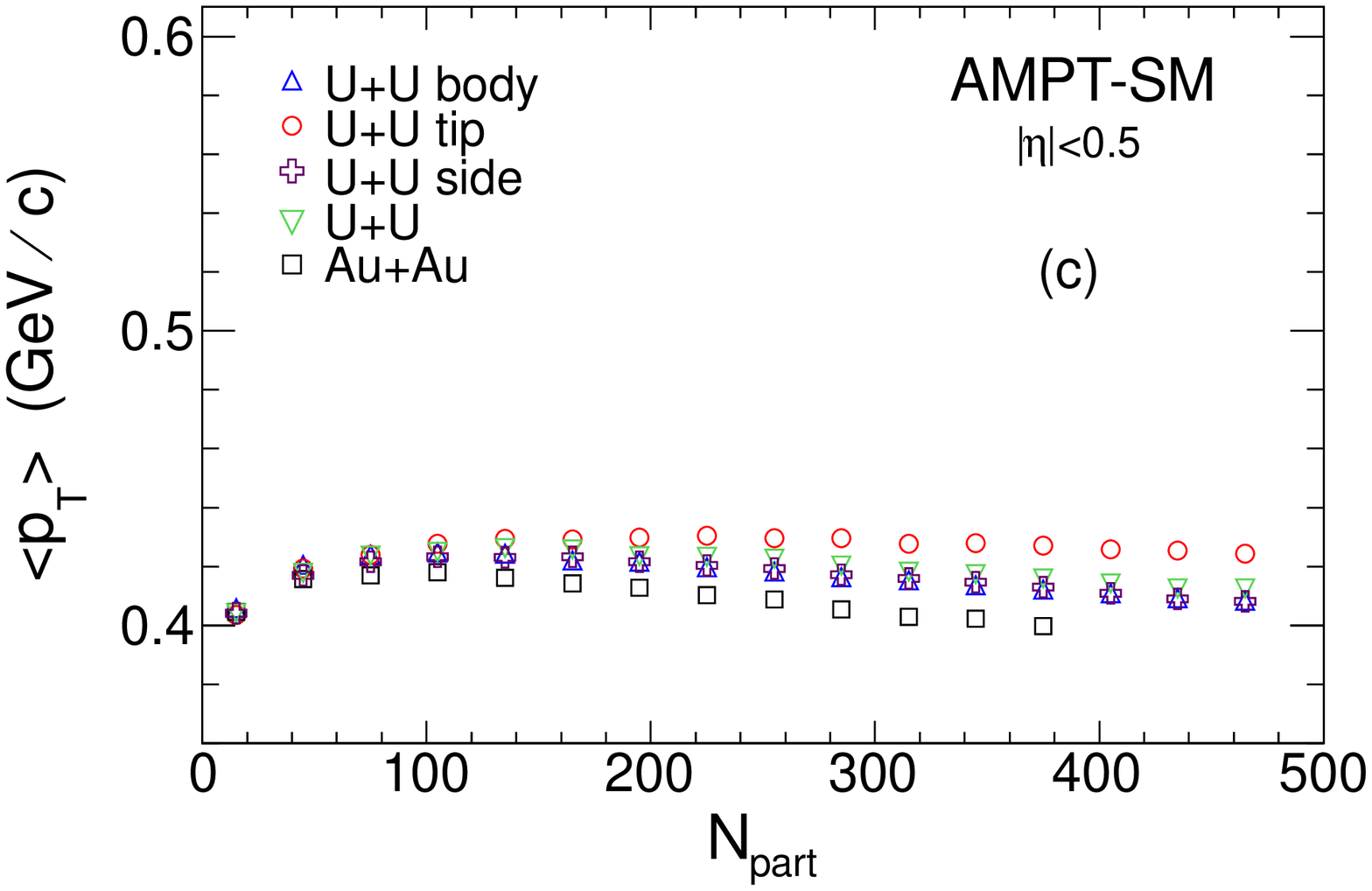}
\caption{(Color online) Same as Fig.~\ref{fig4a} for AMPT string
  melting version.}
\label{fig4b}
\eef

Figure~\ref{fig4a}(a) shows $dN_{\rm {ch}}/d{\eta}$ for central collisions (impact parameter b $<$ 3.6 fm)
vs. $\eta$, Fig.~\ref{fig4a}(b) $(dN_{\rm {ch}}/d{\eta})/(N_{\rm {part}}/2)$ vs. $N_{\rm {part}}$
and Fig.~\ref{fig4a}c charged particle average transverse momentum $\langle p_{\rm T} \rangle$ vs. $N_{\rm {part}}$ for
different collision configuration of U+U collisions and Au+Au collisions
at $\sqrt{s_{\rm {NN}}}$ = 200 from default AMPT model, where $N_{\rm {part}}$ is the number of participating
nucleons. The shape of the $dN_{\rm {ch}}/d{\eta}$ are similar 
for all collision configuration studied and in terms of
multiplicity the conclusions are same as seen in Fig.~\ref{fig3a}(a). 
The $(dN_{\rm {ch}}/d{\eta})/(N_{\rm {part}}/2)$  at at midrapidity 
($\mid \eta \mid$ $<$ 0.5) in U+U collisions extends
to higher $N_{\rm {part}}$ values compared to Au+Au collisions. As expected
it increases with increase in $N_{\rm {part}}$. 
The charged particle multiplicity in most central collisions studied
shows a trend of body-to-body and side-on-side values being similar and lower 
than the values for tip-to-tip case with $(dN_{\rm {ch}}/d{\eta})/(N_{\rm {part}}/2)$ values 
for the general U+U collisions lying in between. The Au+Au collision values for central collisions 
are similar to those from the general U+U configuration case for similar $N_{\rm {part}}$. The charged particle $\langle p_{\rm T} \rangle$ 
at midrapidity increases with increase in $N_{\rm {part}}$. For central collisions
the  $\langle p_{\rm T} \rangle$ for tip-to-tip is about 30 MeV higher than
the body-to-body case with general U+U configuration $\langle p_{\rm T} \rangle$
values lying in between. The increase in $\langle p_{\rm T} \rangle$ 
at midrapidity ($\mid \eta \mid < 0.5$) for U+U tip-to-tip collisions relative
to Au+Au collisions is small and is about 10 MeV.

Figure~\ref{fig4b} shows the same results as in Fig.~\ref{fig4a} using the string melting version
of the AMPT model. The conclusions from $dN_{\rm {ch}}/d{\eta}$ are similar as for the default
case, except that the $dN_{\rm {ch}}/d{\eta}$ values are higher in the string melting version.
The charged particle $\langle p_{\rm T} \rangle$ trends with respect to $N_{\rm {part}}$ is 
however different.  The $\langle p_{\rm T} \rangle$ values are lower for the string melting 
version compared to default case and it saturates or slightly decreases as one goes to
central collisions. The saturation of $\langle p_{\rm T} \rangle$ values for central collisions
in the string melting version could be due to additional partonic interactions and the quark coalescence process 
in the model relative to that for the default case.

\subsection{Geometrical variables}
We have followed the notations for the participant eccentricity ($\ecc$) and 
triangularity ($\tria$) as studied in Ref~\cite{alver}. 
The participant eccentricity is defined as:
\begin{equation}
  \ecc = \frac{\sqrt{\mean{r^2\cos(2\phi_{\text{part}})}^2 + \mean{r^2\sin(2\phi_{\text{part}})}^2}}{\mean{r^2}}
\label{eq:ecc}
\end{equation}
where $r$ and $\phi_{\text{part}}$ are the polar coordinate positions of participating nucleons in the
AMPT model.
Similar to the definition of the eccentricity the
participant triangularity, $\tria$ is defined as:
\begin{eqnarray}
  \tria &=& \frac{\sqrt{\mean{r^2\cos(3\phi_{\text{part}})}^2 + \mean{r^2\sin(3\phi_{\text{part}})}^2}}{\mean{r^2}} \label{eq:tria} 
\end{eqnarray}

\bef
\includegraphics[scale=0.4]{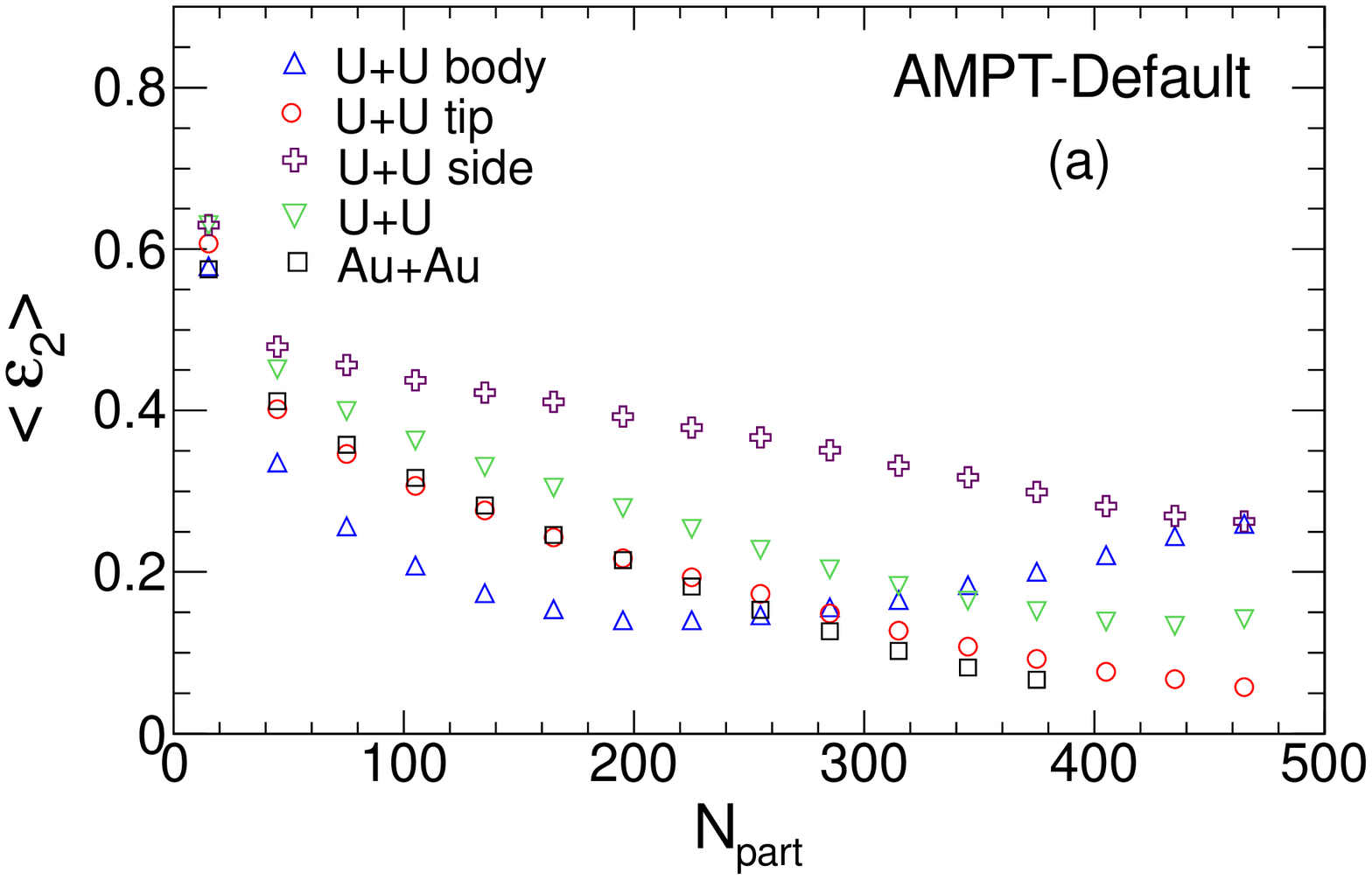}
\includegraphics[scale=0.4]{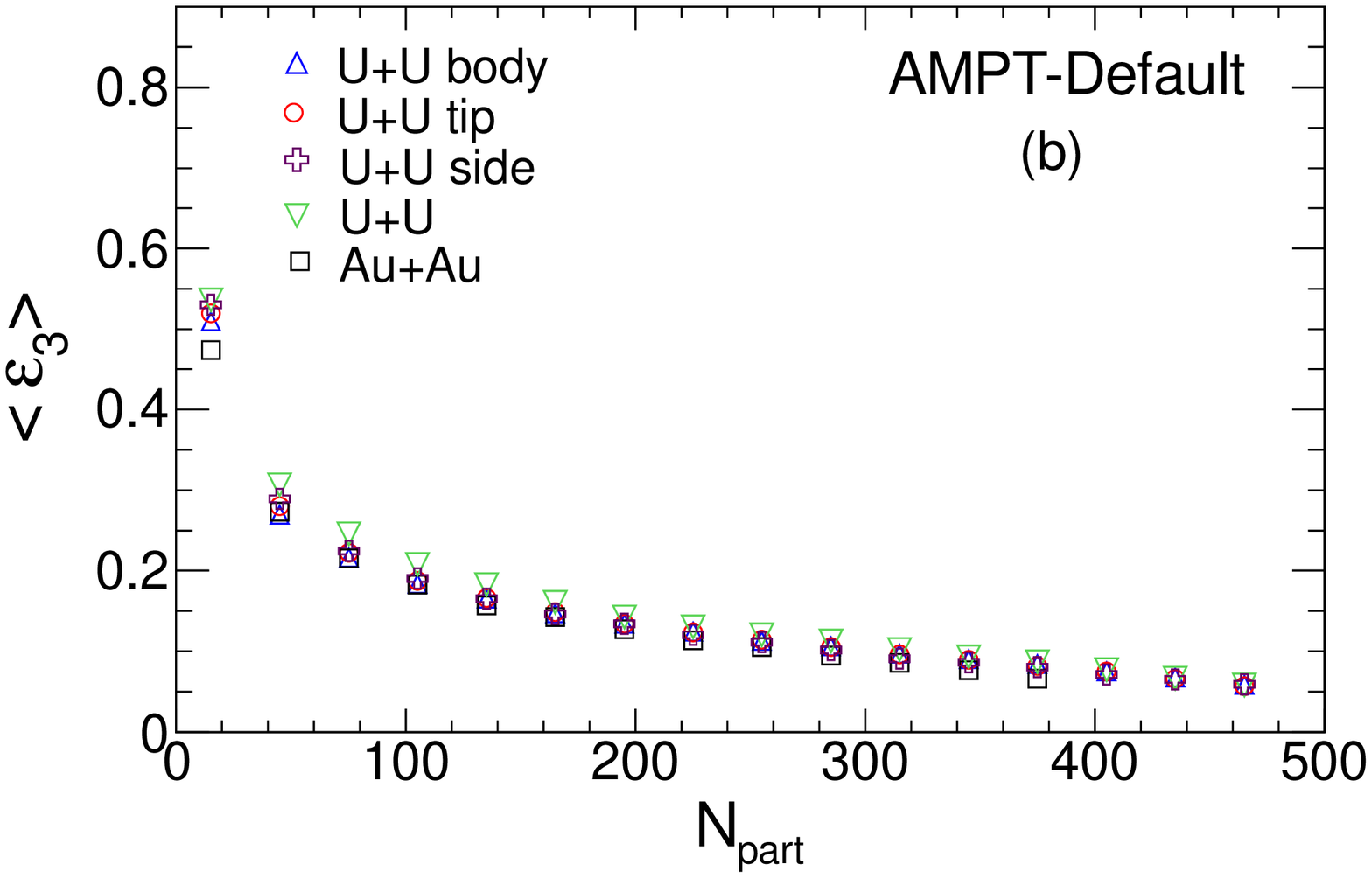}
\caption{(Color online) (a) Participant eccentricity ($\langle \ecc \rangle$) and 
(b) triangularity ($\langle \tria \rangle$) as a 
function of number of participating nucleons ($N_{\rm {part}}$) for various configurations
of U+U collisions and Au+Au collisions at $\sqrt{s_{\rm {NN}}}$ = 200 GeV from default AMPT model. }
\label{fig5}
\eef

Figure~\ref{fig5} shows the $\langle \ecc \rangle$ and $\langle \tria \rangle$ vs. 
$N_{\rm {part}}$ for various configurations
of U+U collisions at $\sqrt{s_{\rm {NN}}}$ = 200 GeV and Au+Au collisions at 
$\sqrt{s_{\rm {NN}}}$ = 200 GeV from default AMPT model. For the same $N_{\rm {part}}$ the U+U collisions
without any specific selection of collision configuration have higher $\langle \ecc \rangle$ compared 
to Au+Au collisions. For U+U collisions the tip-to-tip configuration has a lower $\langle \ecc \rangle$ compared
to no specific selection of collision configuration. The side-on-side  configuration 
have the largest values of  $\langle \ecc \rangle$ for the systems studied. The $\langle \ecc \rangle$
for body-to-body configuration shows a specific trend as a function of $N_{\rm {part}}$, it is
similar to side-on-side and tip-to-tip in most peripheral collisions, then decreases sharply
to values below those from tip-to-tip collisions for mid central collisions which is followed
by an increase in values of $\langle \ecc \rangle$ with  $N_{\rm {part}}$ to reach the same
values as side-on-side  for the most central collisions. This clearly reflects the 
specific geometrical configuration traversed by the two Uranium nuclei in different cases. 
The $\langle \tria \rangle$ however is found to be similar for all configurations 
in U+U studied and for Au+Au collisions as a function of $N_{\rm {part}}$.
Since these are specific to geometrical configurations of the nuclei, we observe no
difference in these variable for the string melting version of the model.

Next we study the fluctuation in $\ecc$ and $\tria$ as it has important consequences on
understanding of the initial conditions in heavy-ion collisions as well as flow fluctuations.
The observables used are the ratio of root mean square (rms) value of $\ecc$ 
to $\langle \ecc \rangle$, rms of $\tria$ to $\langle \tria \rangle$ and those suggested
in Ref.~\cite{bhalerao}: $\langle\varepsilon_n^4\rangle/\langle\varepsilon_n^2\rangle^2$
(for $n=2,3$ in this work). 
\bef
\includegraphics[scale=0.35]{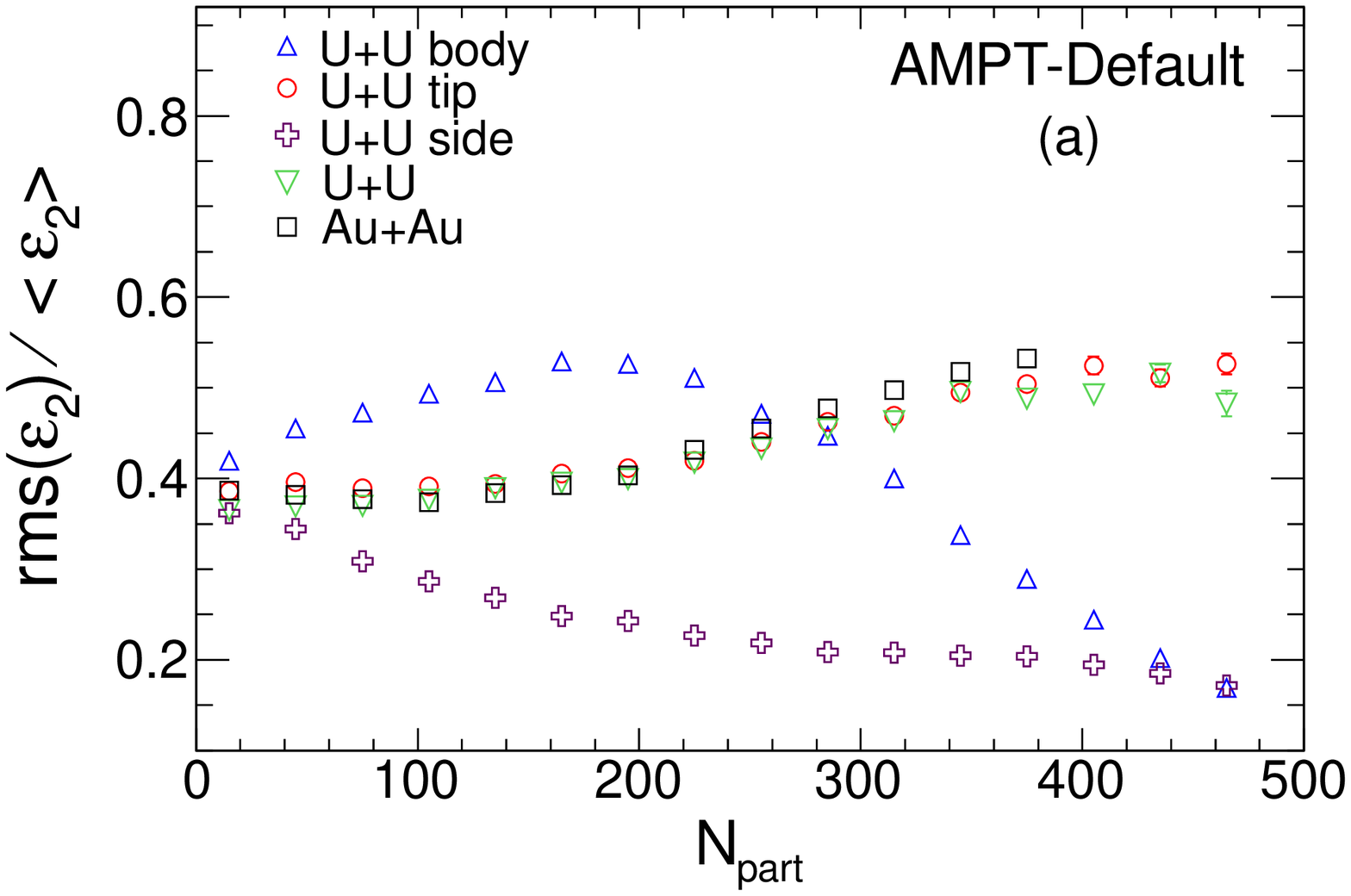}
\includegraphics[scale=0.35]{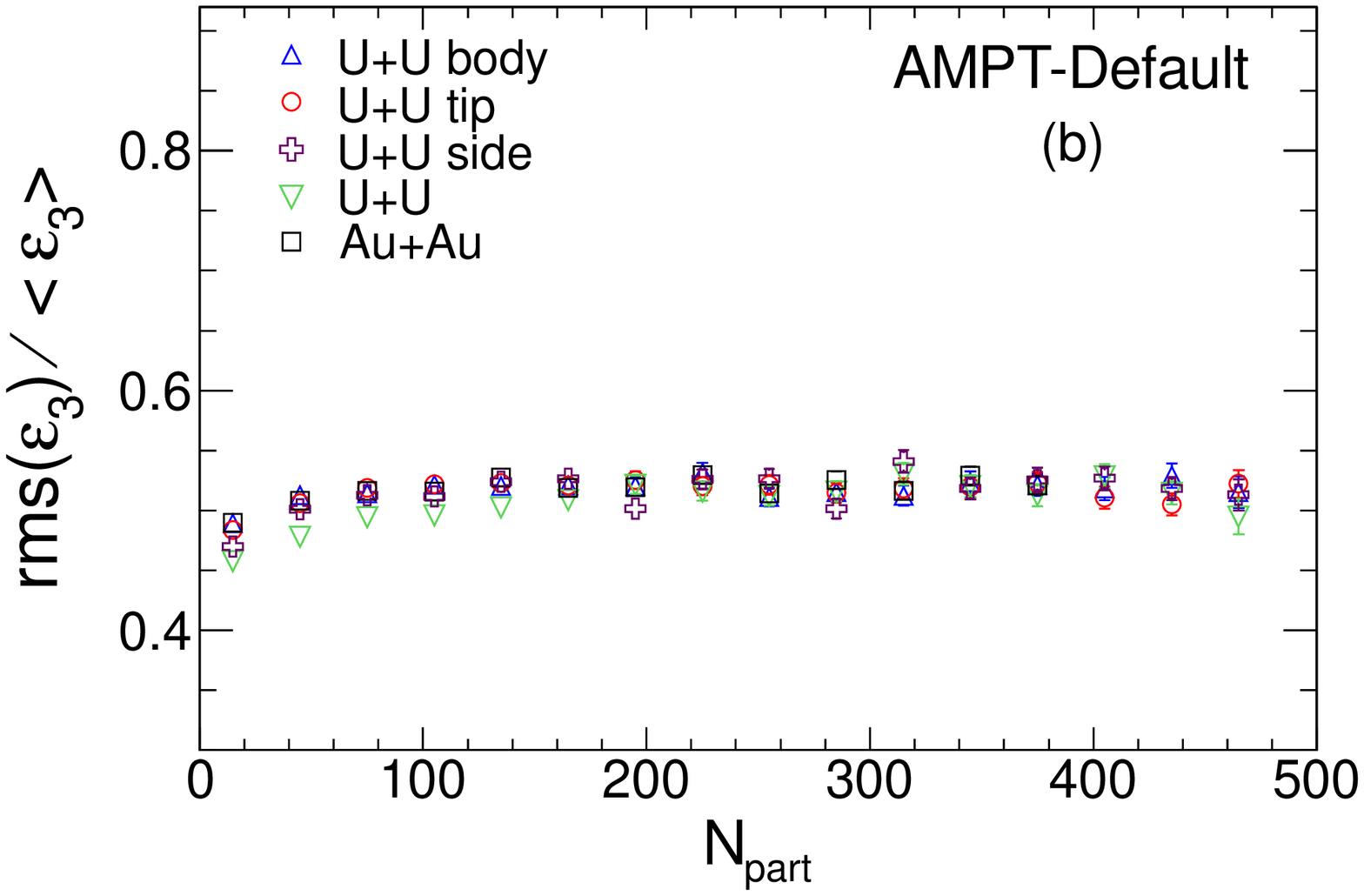}
\includegraphics[scale=0.35]{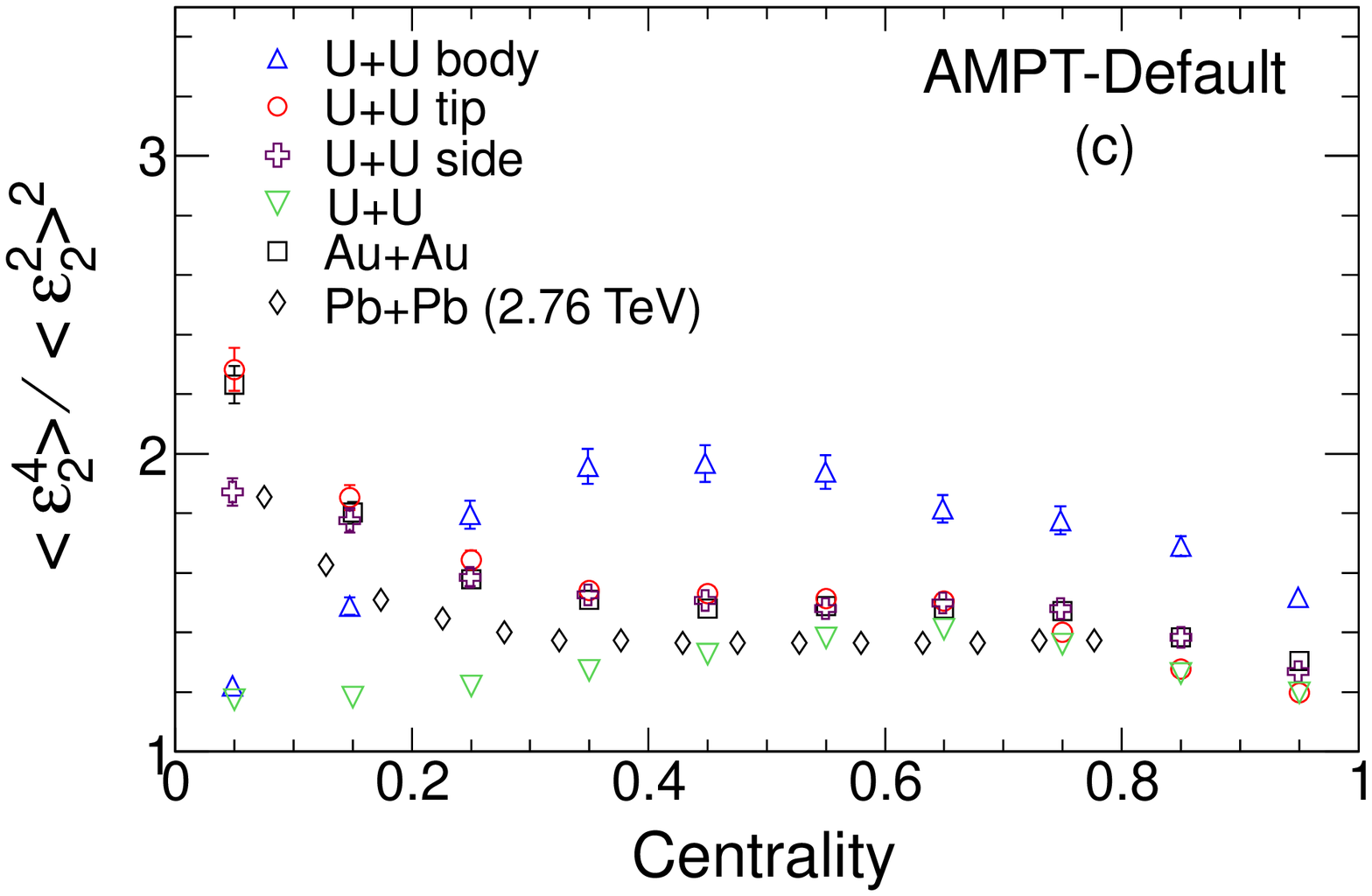}
\includegraphics[scale=0.35]{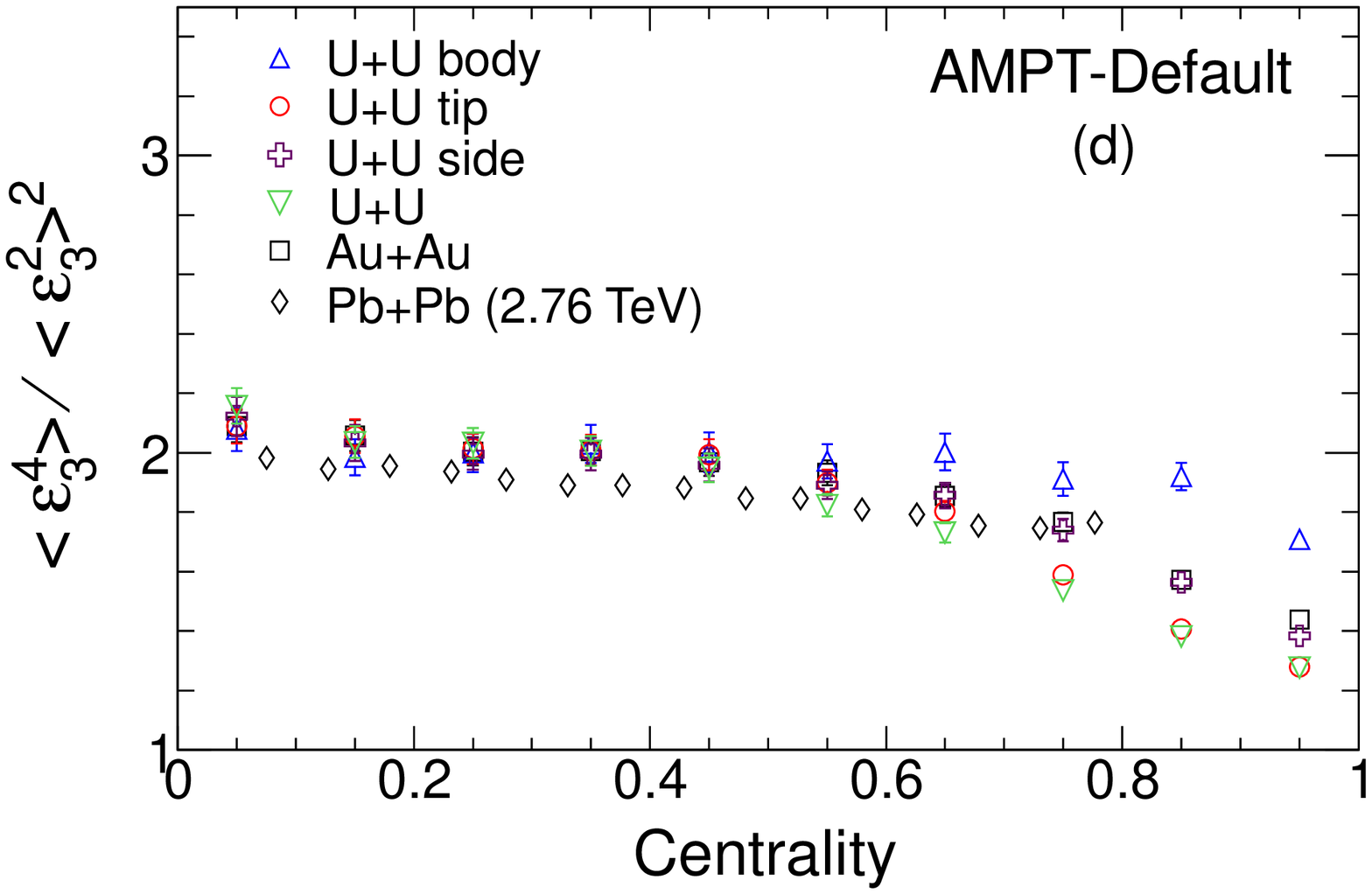}
\caption{(Color online) (a) Ratio of root mean square (rms) value of $\ecc$ 
to $\langle \ecc \rangle$ and (b) rms of $\tria$ to $\langle \tria \rangle$ vs. $N_{\rm {part}}$ 
for various configurations of U+U collisions and Au+Au collisions at $\sqrt{s_{\rm {NN}}}$ = 200 GeV 
using default AMPT model. (c) and (d) $\langle\varepsilon_n^4\rangle/\langle\varepsilon_n^2\rangle^2$, 
with $n=2,3$, versus fraction of collision centrality for U+U and 
Au+Au collisions at $\sqrt{s_{\rm {NN}}}$ = 200 GeV
using the default AMPT model. The Pb+Pb results corresponds to Glauber model simulations 
from Ref.~\cite{bhalerao} at $\sqrt{s_{\rm {NN}}}$ = 2.76 TeV.}
\label{fig6}
\eef
Figure~\ref{fig6}(a) and (b) shows the event-by-event fluctuations in 
$\ecc$ and $\tria$ as a function of $N_{\rm {part}}$ for U+U and Au+Au collisions at
$\sqrt{s_{\rm {NN}}}$ = 200 GeV respectively. The fluctuations in $\ecc$ for U+U collisions with
no specific selection on collision configuration closely follows those for Au+Au collisions,
however for the most central collisions the fluctuations are slightly smaller for U+U collisions.
The fluctuations in $\ecc$ for tip-to-tip configuration are comparable to those for Au+Au collisions.
The fluctuations in $\ecc$ for side-on-side configuration are the smallest among the 
configurations studied. On the other hand, those for body-to-body U+U collisions reflects an unique trend with
fluctuations in $\ecc$ being largest in mid-central collisions and then decreasing with increase
in centrality to reach the corresponding values of side-on-side for central most collisions.
Exactly similar trends are observed using the variable 
$\langle\varepsilon_n^4\rangle/\langle\varepsilon_n^2\rangle^2$ as a function of fraction of collision centrality 
(Fig.~\ref{fig6} (c) and (d)). In the Figs.~\ref{fig6} (c) and (d)  the x-axis value near 0 means most-peripheral 
and the value near 1 means most-central collisions. The centrality is determined from the impact parameter distribution.
The fluctuation in  $\tria$ are observed to be independent of the collision configuration
in U+U and similar to Au+Au collisions, except perhaps for the central most collisions.
For comparison results from a Glauber model simulation for Pb+Pb collisions at
$\sqrt{s_{\rm {NN}}}$ = 2.76 TeV from Ref.~\cite{bhalerao} are also shown.
Our study shows that if different U+U configurations can be selected in experimental data, it 
would lead to interesting variations of flow and flow fluctuations as a function of 
collision centrality, thereby providing a way to understand initial conditions in heavy-ion
collisions at high energies.

\subsection{Elliptic and Triangular flow}

\bef
\includegraphics[scale=0.4]{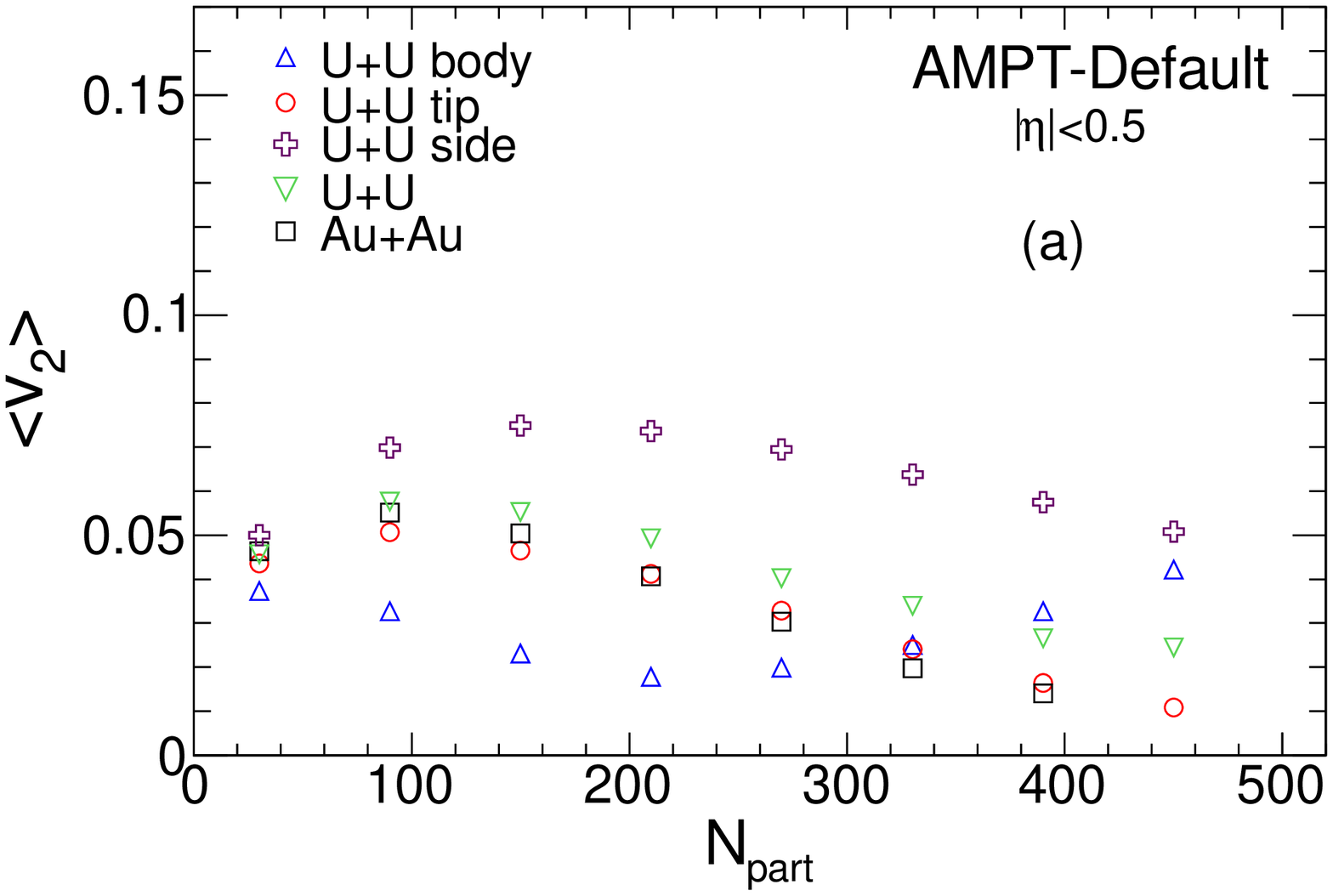}
\includegraphics[scale=0.4]{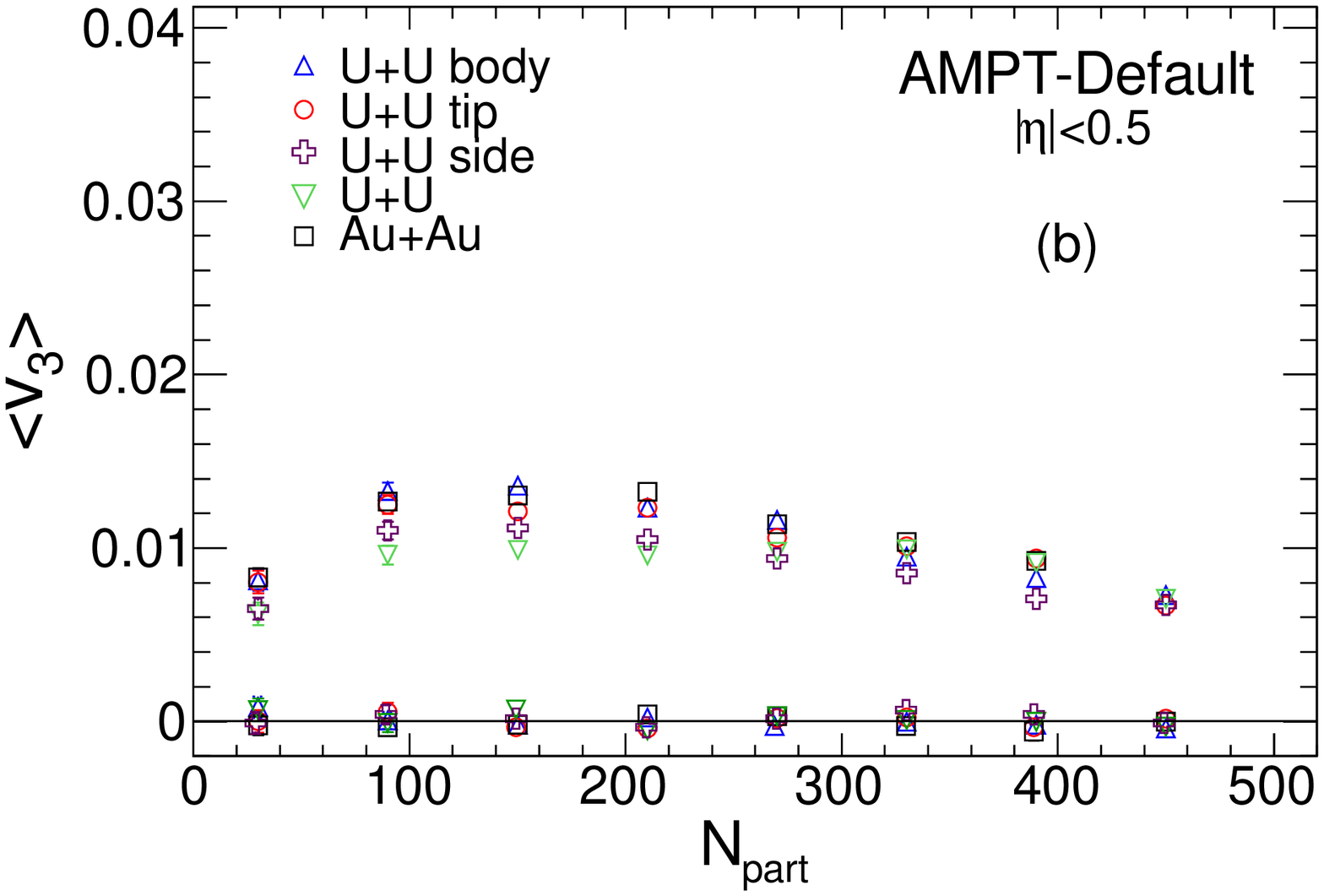}
\caption{(Color online) (a) Average elliptic flow ($\langle v_{\rm 2} \rangle$) and 
(b) triangular flow ($\langle v_{\rm 3} \rangle$) versus $N_{\rm {part}}$ 
 for different collision configuration 
of U+U and Au+Au collisions at midrapidity for $\sqrt{s_{\rm {NN}}}$ = 200 GeV from default AMPT model. 
In (b) the $\langle v_{\rm 3} \rangle$ values close to zero are those corresponding to $\langle v_{\rm 3} \rangle$
calculated using $\psi_{2}$.}
\label{fig7a}
\eef

\bef
\includegraphics[scale=0.4]{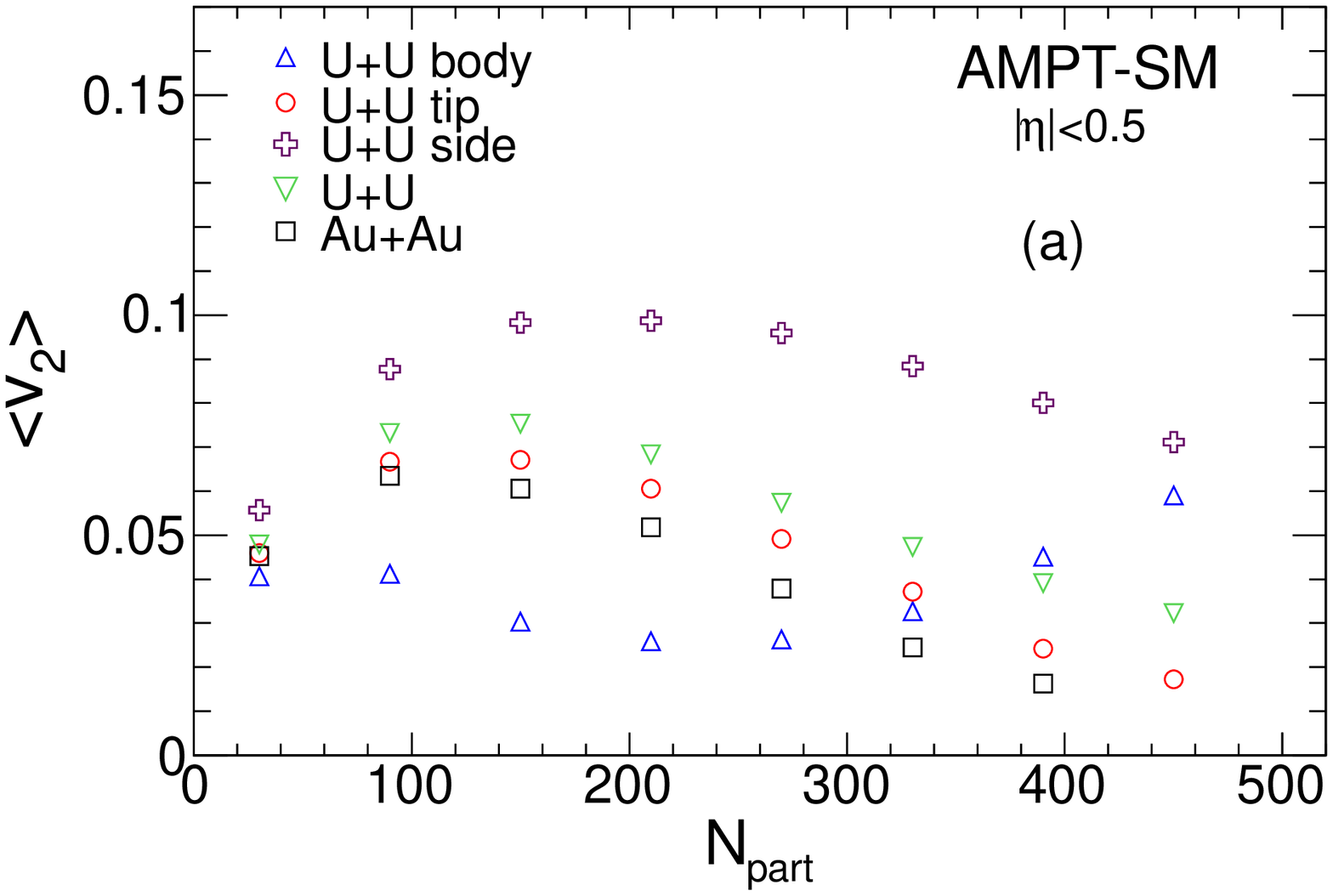}
\includegraphics[scale=0.4]{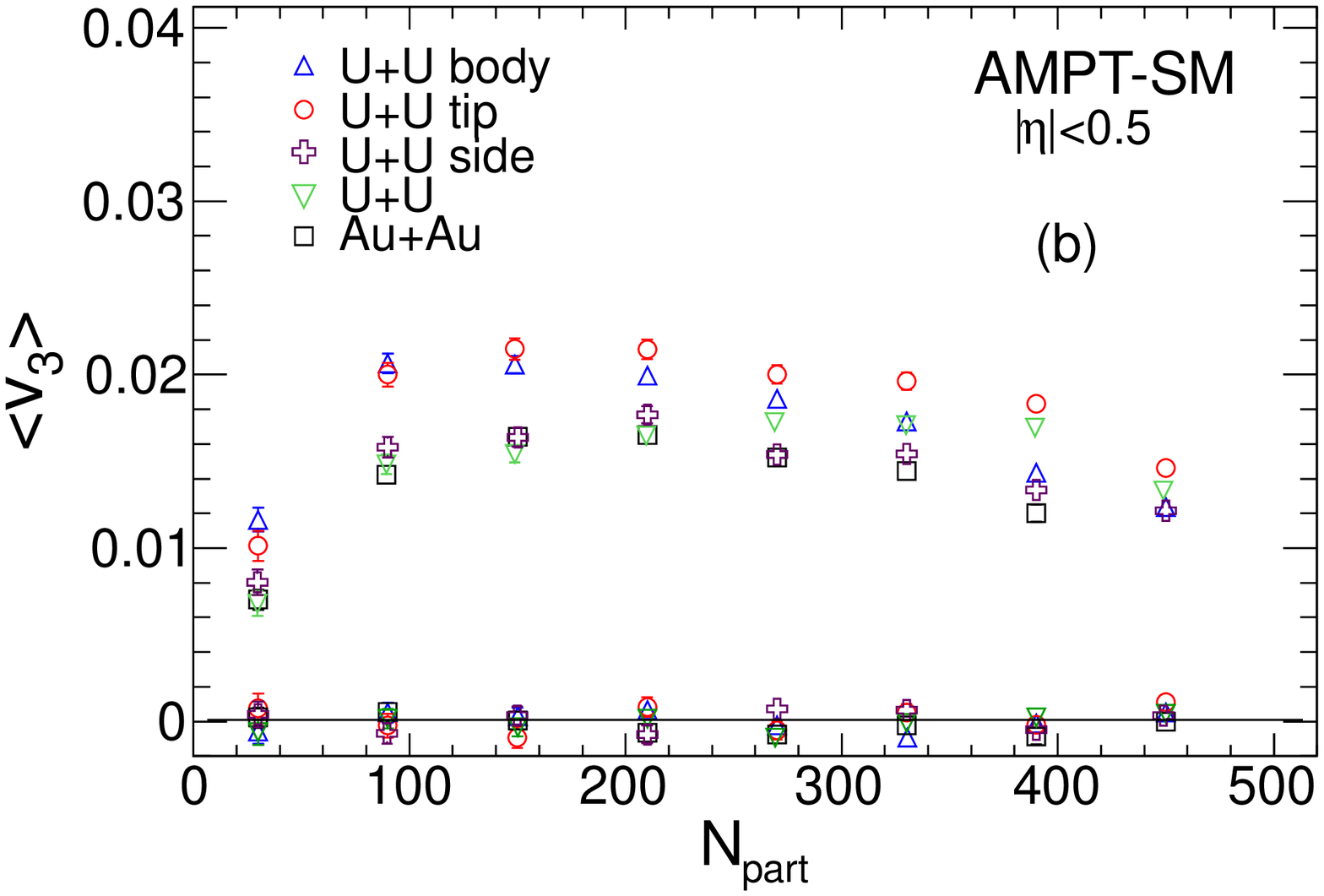}
\caption{(Color online) Same as Fig.~\ref{fig7a} for AMPT string melting version.}
\label{fig7b}
\eef

The elliptic flow $v_2$ which is the second Fourier coefficient of particle distribution 
with respect to $\psi_{2}$ is given as 
\begin{equation}
v_2 = \mean{\cos(2(\phi-\psi_2))}
\label{eq:v2}
\end{equation}
where  $\psi_{2}$ is the minor axis of the ellipse defined as
\begin{equation}
  \psi_{2}=\frac{\atantwo\left(\mean{r^2\sin(2\phi_{\text{part}})},\mean{r^2\cos(2\phi_{\text{part}})}\right)
+\pi}{2}.
\label{eq:phiecc}
\end{equation}

Similar to the definition of the elliptic flow and $\psi_{2}$, the
triangular flow, $v_3$ and $\psi_{3}$ are defined as:
\begin{equation}
v_3 = \mean{\cos(3(\phi-\psi_3))}
\label{eq:v3}
\end{equation}
 where $\psi_3$ is the minor axis of participant triangularity given by
 \begin{equation}
   \psi_{3}=\frac{\atantwo\left(\mean{r^2\sin(3\phi_{\text{part}})},\mean{r^2\cos(3\phi_{\text{part}})}\right)
     +\pi}{3}.
\label{eq:phitria}
\end{equation}

Figure~\ref{fig7a}(a) and (b) shows the $\langle v_{\rm 2} \rangle$ and $\langle v_{\rm 3} \rangle$
as a function of $N_{\rm {part}}$ at midrapidity ($\mid \eta \mid < 0.5$) for different
configurations of U+U collisions and Au+Au collisions at $\sqrt{s_{\rm {NN}}}$ = 200 GeV.
The characteristic trend of centrality dependence (smaller values for central collisions 
and larger values for mid-central collisions)  of $\langle v_{\rm 2} \rangle$ is
observed for most of the configurations studied except for U+U body-to-body collisions.
In fact the body-to-body collisions shows a minimum $\langle v_{\rm 2} \rangle$ for mid-central
collisions which is consistent with the variation of $\langle \ecc \rangle$ with centrality
as shown in Fig.~\ref{fig6}(a). Figure~\ref{fig7a}(b) shows the corresponding results
for $v_{\rm 3}$. It is found $v_{3}$ is slightly higher for Au+Au collisions compared to
U+U collisions without any choice of configuration. For U+U collisions with various configurations the 
largest $v_{\rm 3}$ seems to be from body-to-body condition, while those for side-on-side are smaller. 
Also shown in ~\ref{fig7a}(b) are the $\langle v_{\rm 3} \rangle$ values (close to zero) when calculated
using  $\psi_{2}$ instead of  $\psi_{3}$. The $\langle v_{\rm 3} \rangle$ value of zero shows that the 
minor axis of triangularity is found to be uncorrelated with the reaction plane angle for both U+U and Au+Au collisions.
The corresponding results for the string melting version are shown in Fig.~\ref{fig7b}. The conclusions are
same as for the default case, except that the magnitude of the $\langle v_{\rm 2} \rangle$ and
$\langle v_{\rm 3} \rangle$ are typically 40\% and 80\% higher respectively. Also for $\langle v_{\rm 3} \rangle$ the tip-to-tip
configuration have the largest values while those from  Au+Au collisions have smallest values.

\bef
\includegraphics[scale=0.4]{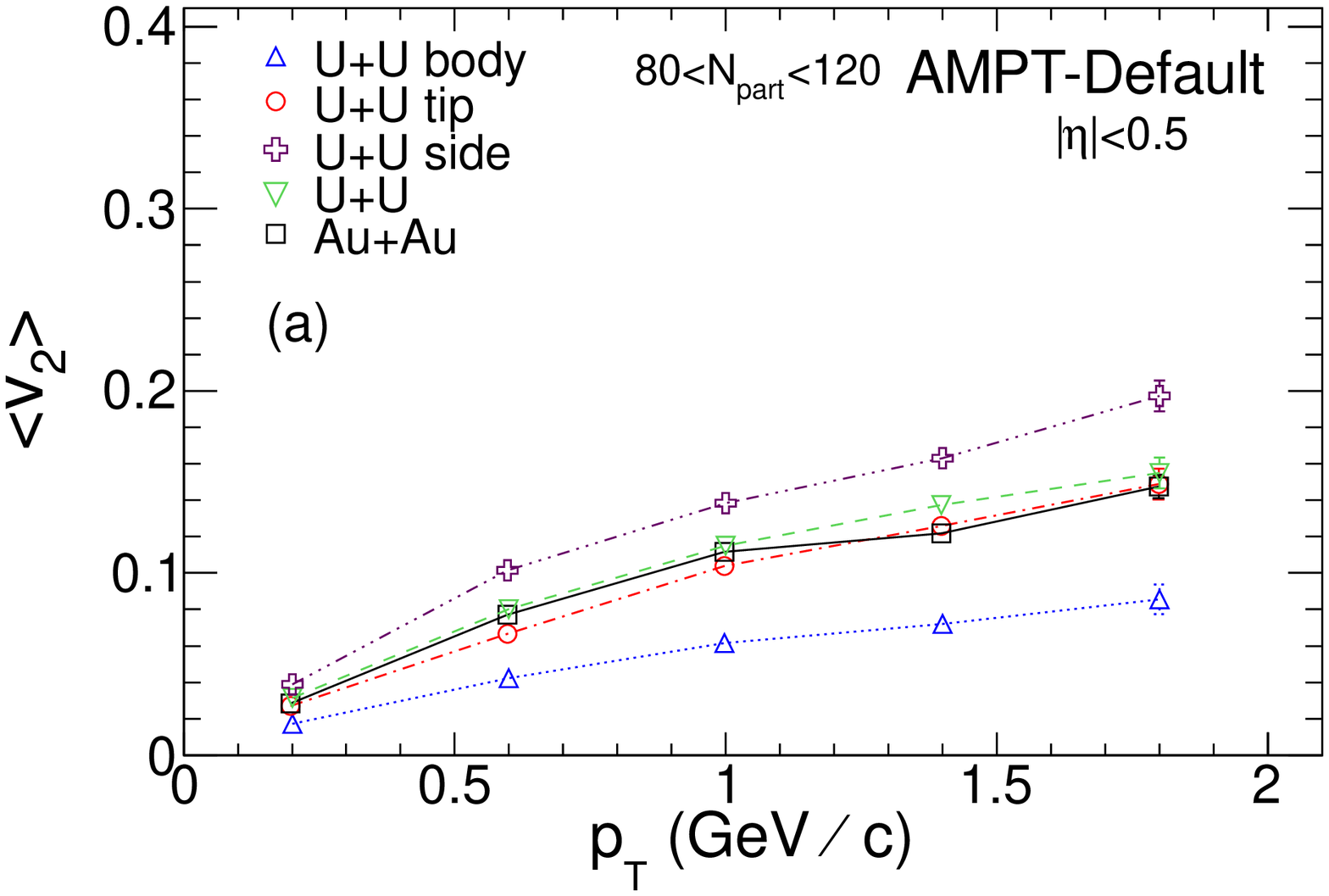}
\includegraphics[scale=0.4]{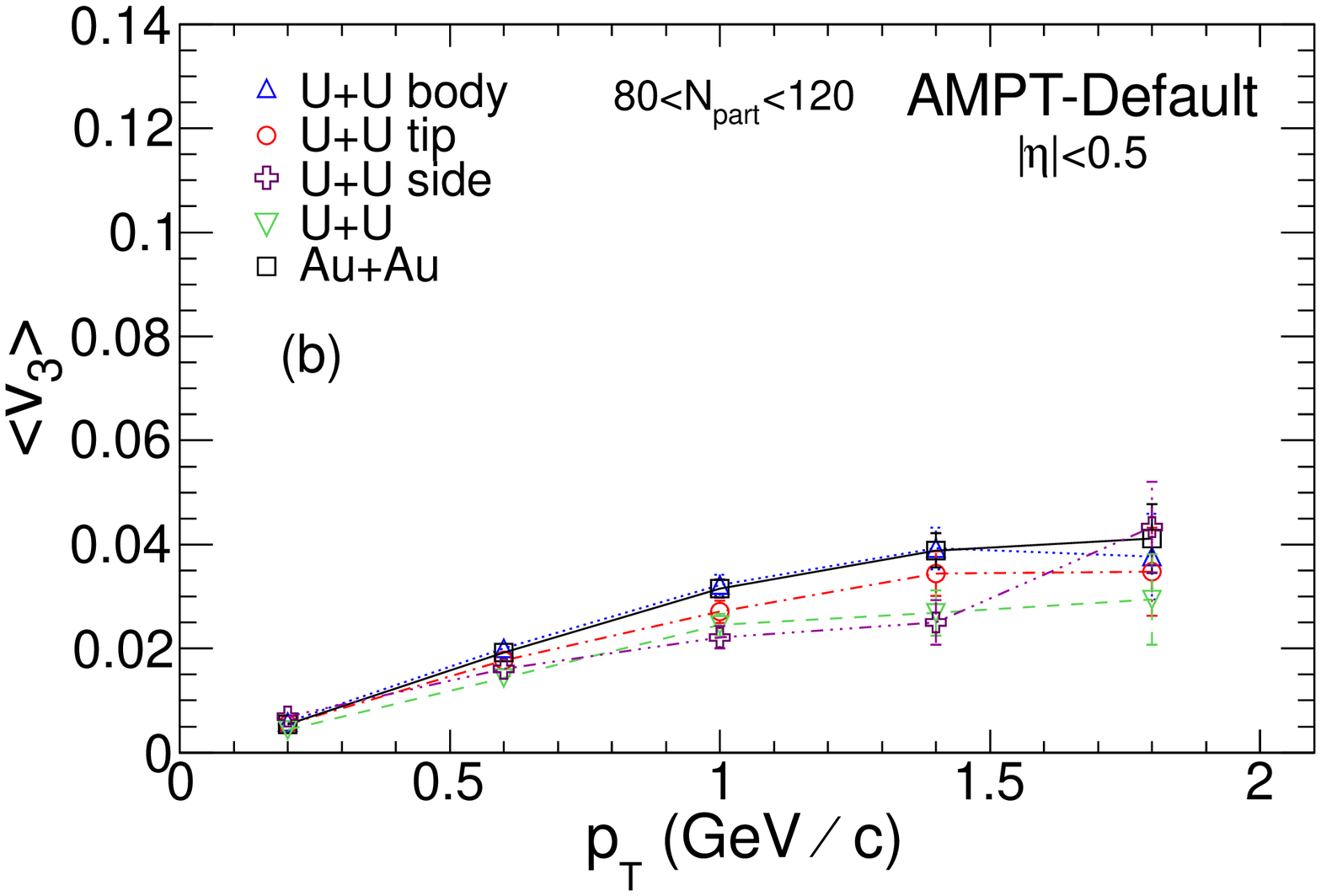}
\caption{(Color online) (a) Elliptic flow ($v_{\rm 2}$) and (b) triangular flow ($v_{3}$) as a 
function of transverse momentum ($p_{\rm T}$) at midrapidity for $80 < N_{\rm {part}} < 120$ 
U+U collisions for different configurations and Au+Au collisions at $\sqrt{s_{\rm {NN}}}$ = 200 GeV from default AMPT
model.}
\label{fig8a}
\eef

\bef
\includegraphics[scale=0.4]{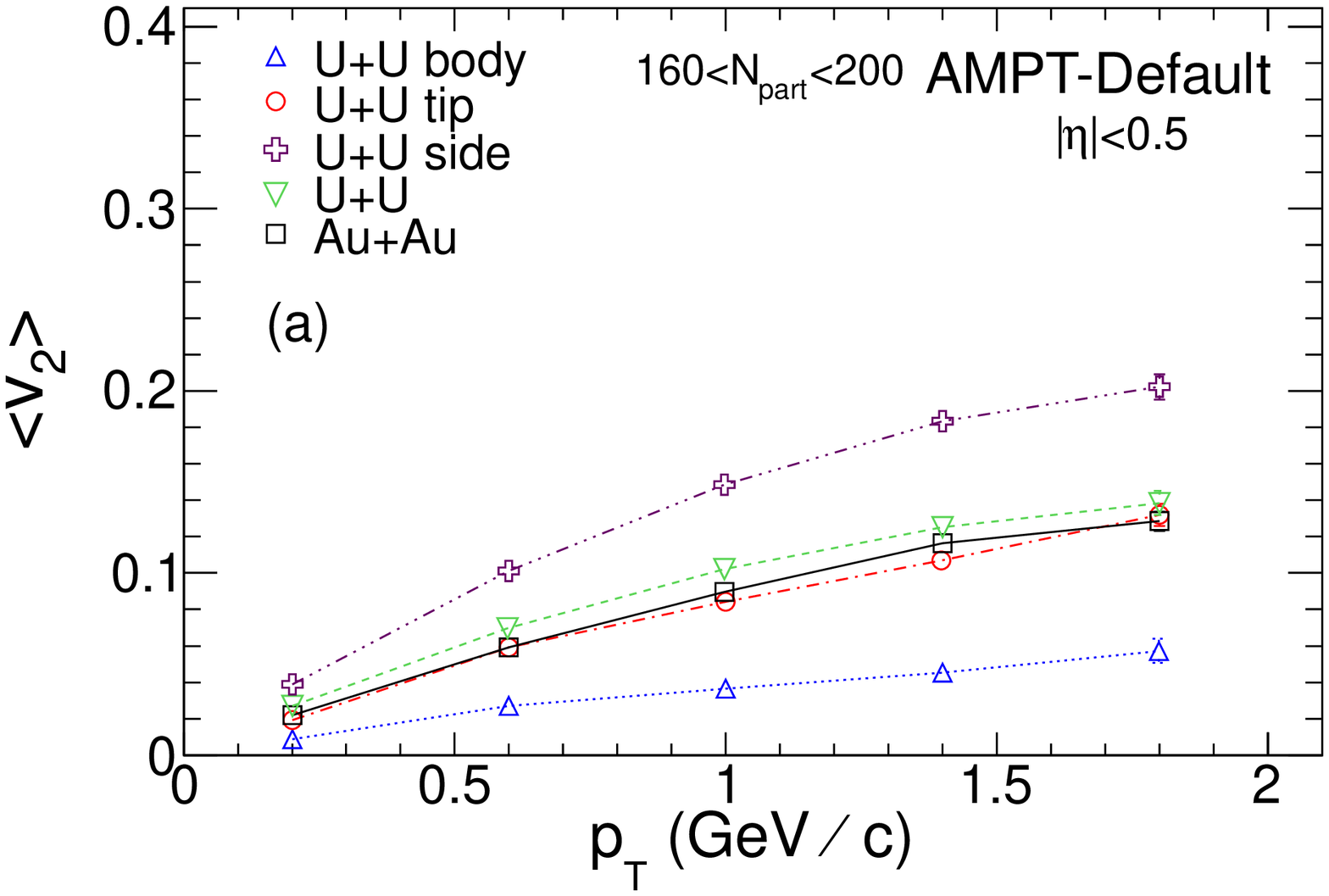}
\includegraphics[scale=0.4]{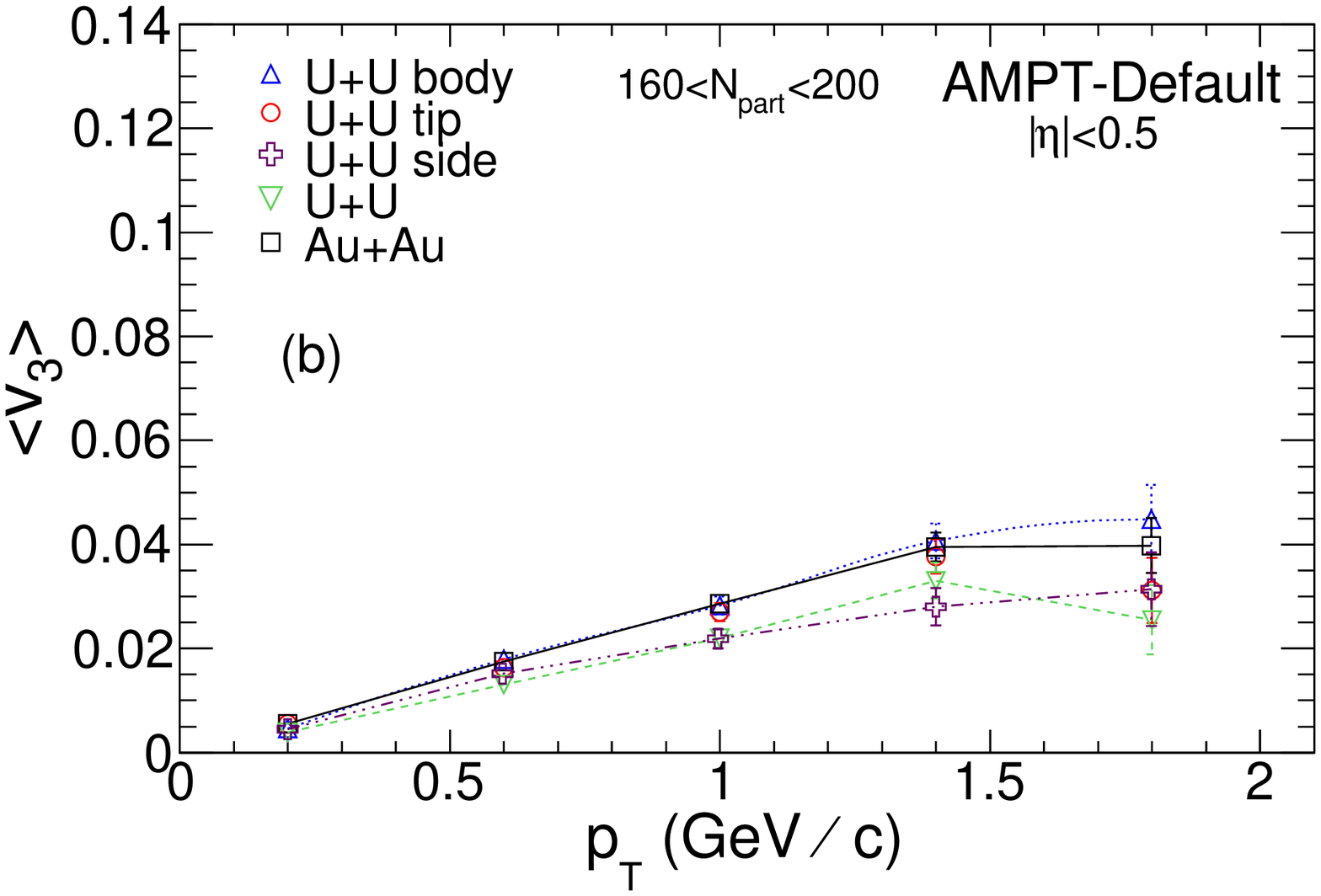}
\caption{(Color online) Same as Fig.~\ref{fig8a} for $160 < N_{\rm {part}} < 200$  .}
\label{fig8aa}
\eef

\bef
\includegraphics[scale=0.4]{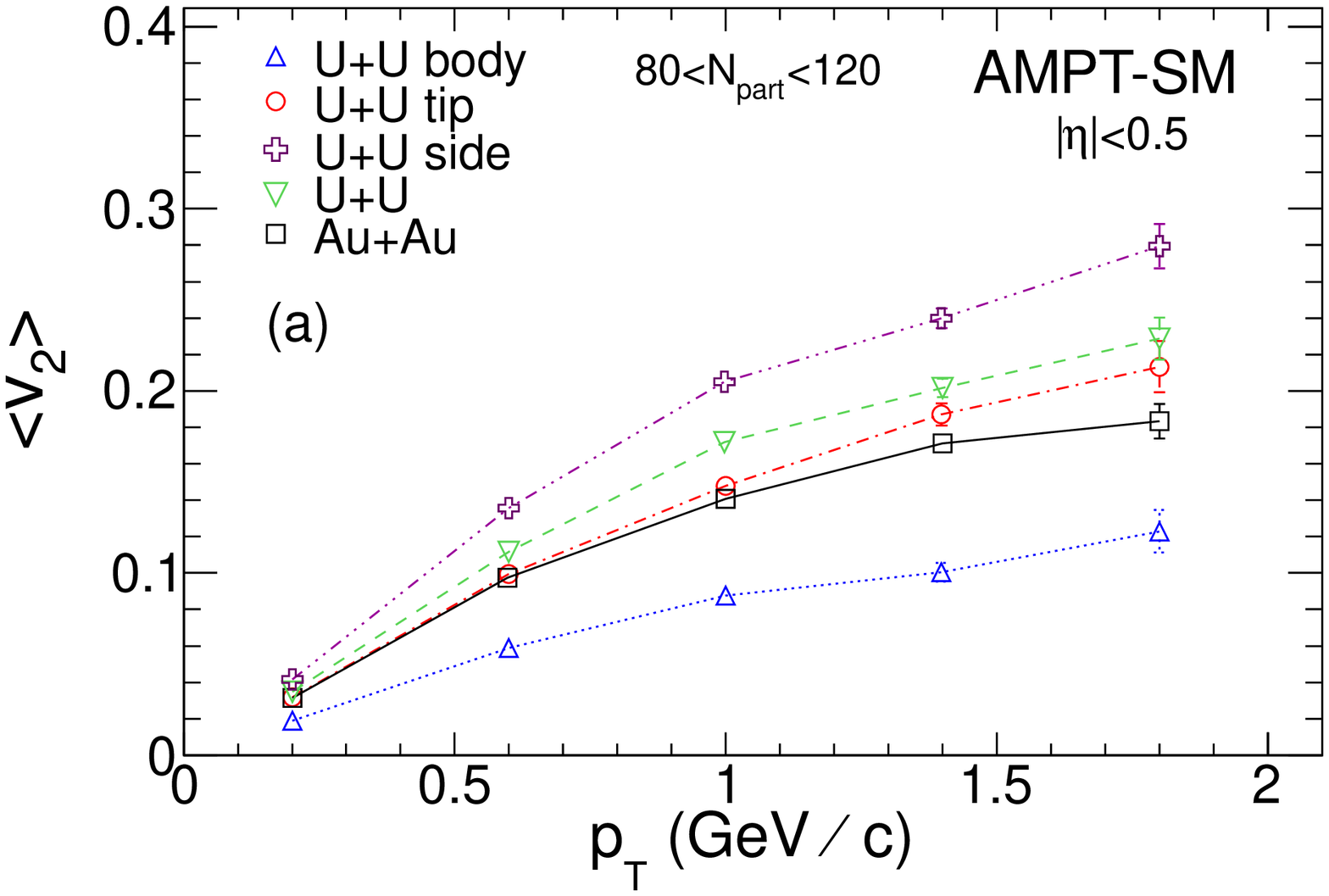}
\includegraphics[scale=0.4]{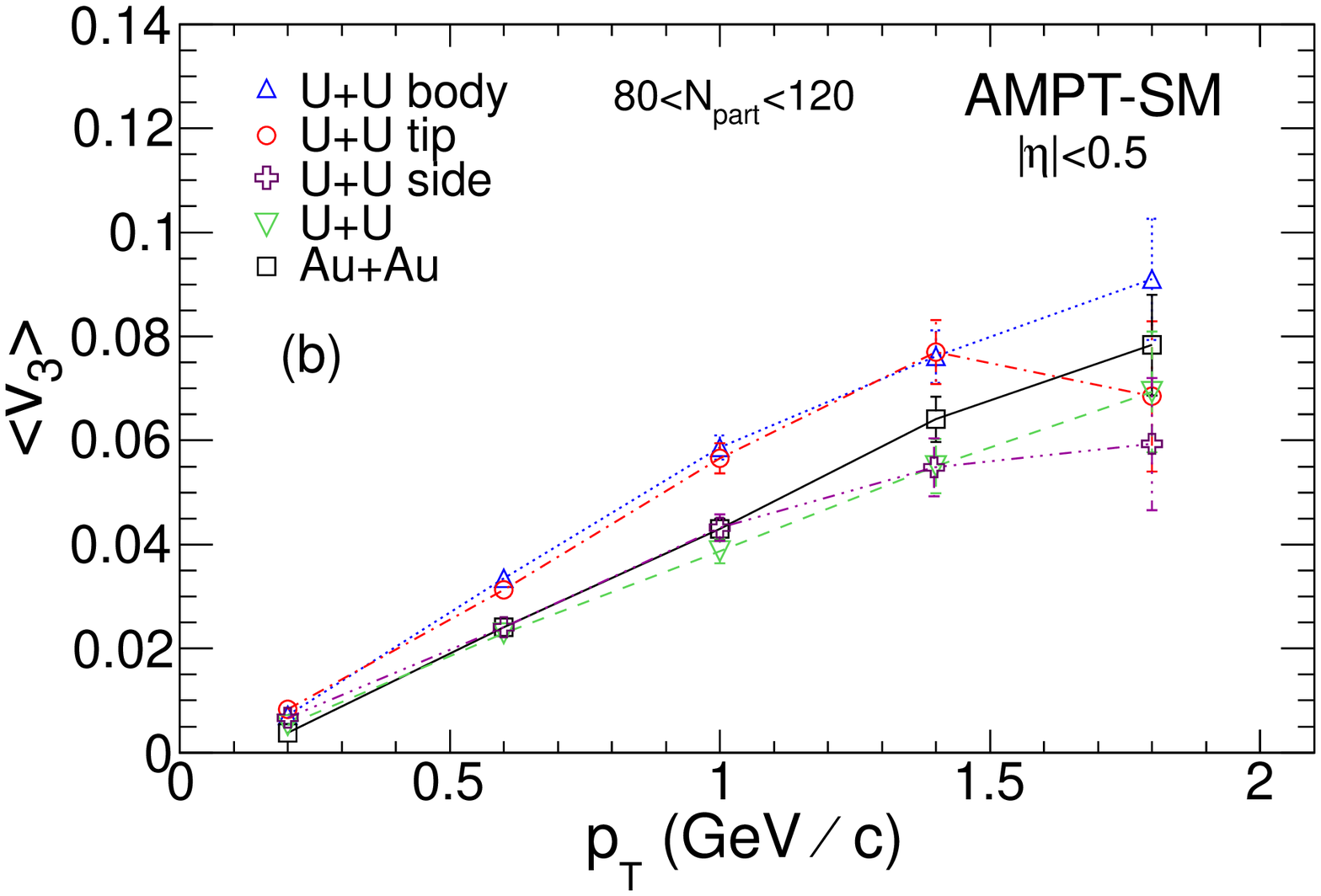}
\caption{(Color online) Same as Fig.~\ref{fig8a} for AMPT string
  melting version.}
\label{fig8b}
\eef

\bef
\includegraphics[scale=0.4]{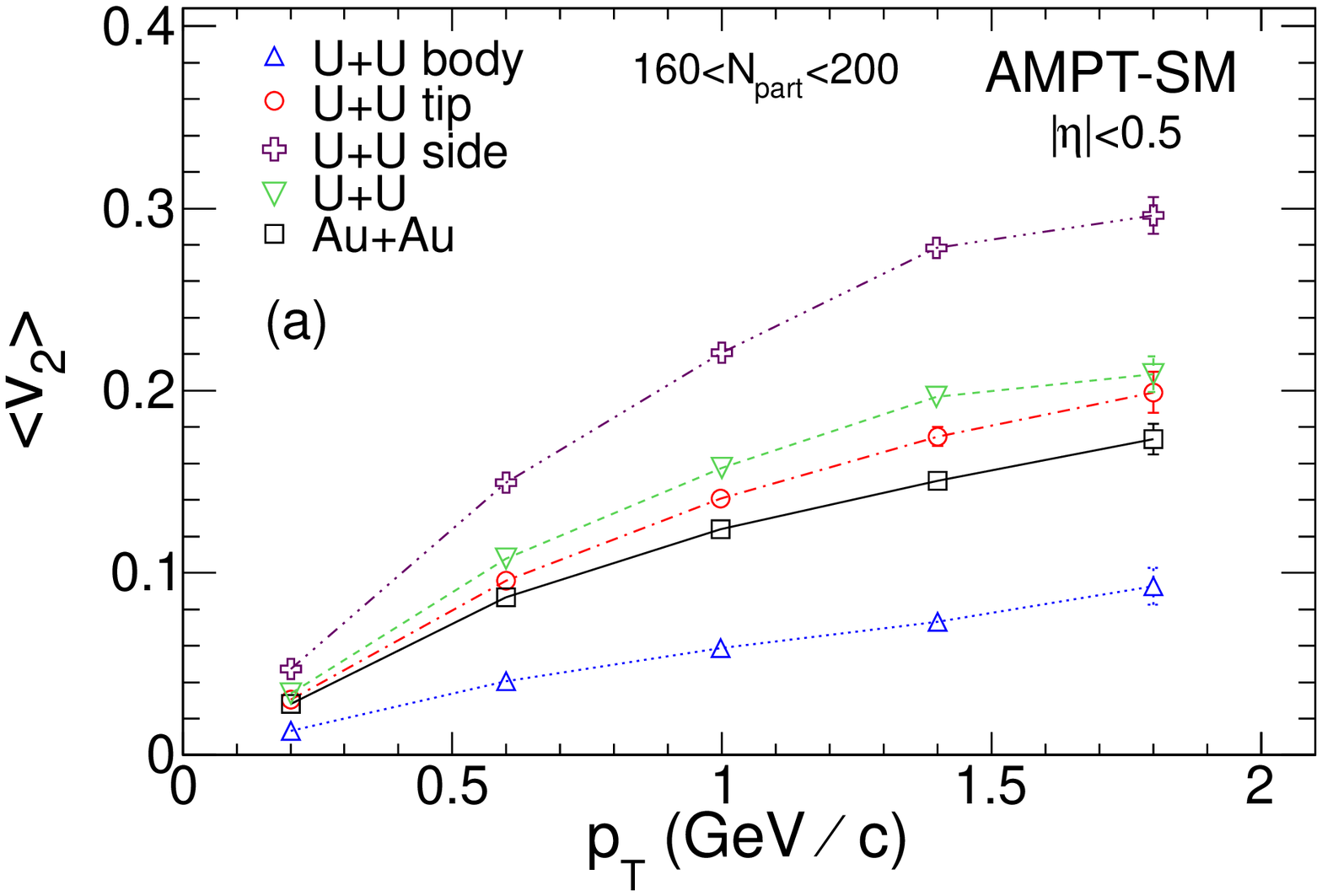}
\includegraphics[scale=0.4]{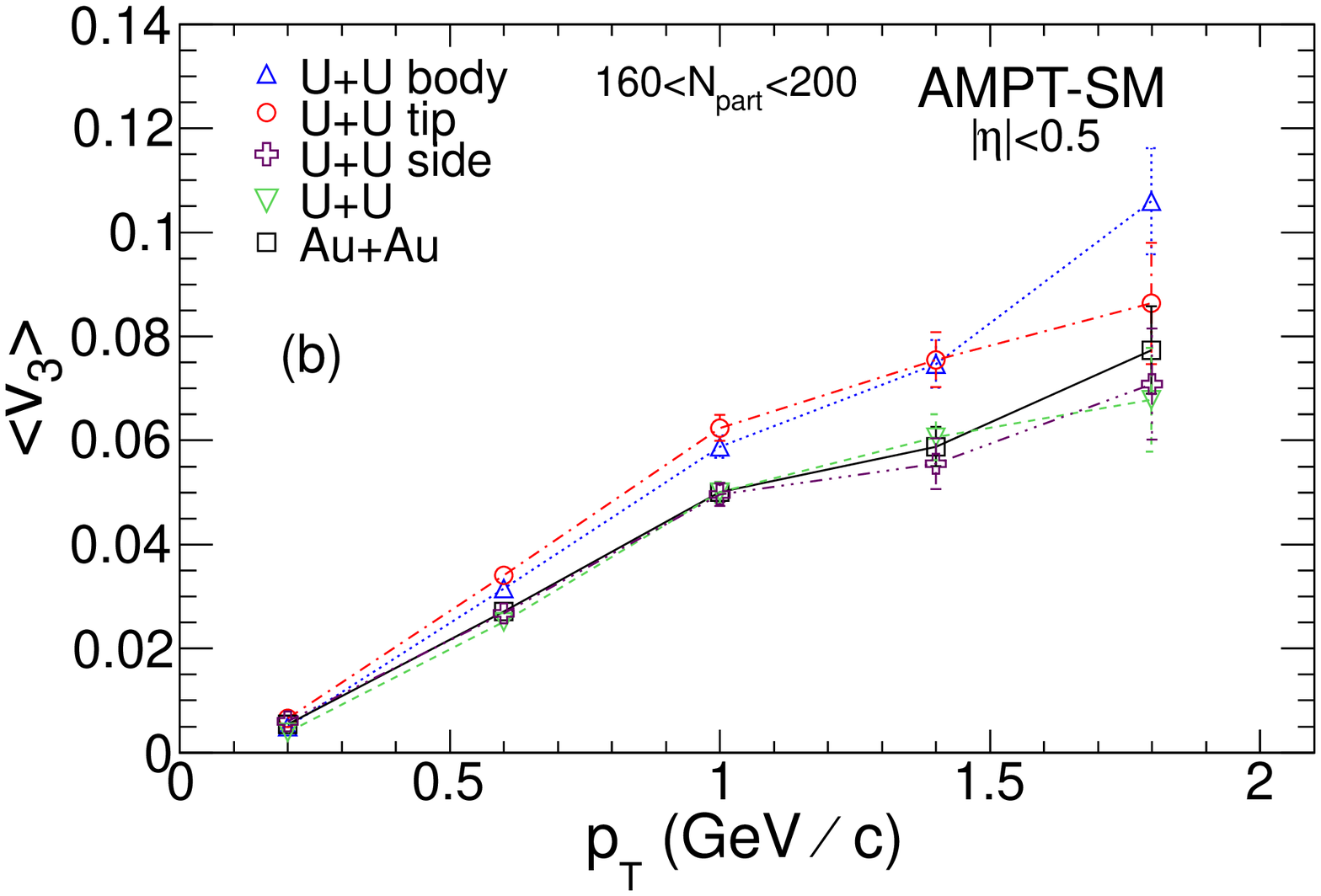}
\caption{(Color online) Same as Fig.~\ref{fig8b} for $160 < N_{\rm {part}} < 200$.}
\label{fig8bb}
\eef

Figure~\ref{fig8a}  and ~\ref{fig8aa} shows the transverse momentum ($p_{\rm T}$) dependence of 
$v_{2}$ and $v_{3}$ for different collision configuration of U+U and Au+Au
collisions at midrapidity ($\mid \eta \mid < 0.5$) 
at $\sqrt{s_{\rm {NN}}}$ = 200 GeV for for $80 < N_{\rm {part}} < 120$
and for $160 < N_{\rm {part}} < 200$ respectively. 
The $v_{2}(p_{\rm T})$ for U+U collisions without any specific collision configuration,
tip-to-tip and Au+Au collisions have similar values for the $p_{\rm T}$ range studied.
The results for $v_{2}(p_{\rm T})$ from body-to-body  U+U collisions are smaller and those
for side-on-side configuration in U+U collisions are higher compared to Au+Au
collisions at similar $p_{\rm T}$ values. The Fig.~\ref{fig8a}(b) shows the
corresponding results for $v_{\rm 3}$. The general trend as observed for $p_{T}$ 
integrated $v_{3}$ (shown in Fig.~\ref{fig7a}(b)) is also followed by $v_{3}(p_{\rm T})$.
The $v_{3}(p_{\rm T})$ for Au+Au and  U+U body-to-body configurations seems to be slightly
higher compared to those from U+U tip-to-tip and U+U with no specific configuration selected.

Figure~\ref{fig8b} and ~\ref{fig8bb} shows the corresponding results as given in
Fig~\ref{fig8a}  and ~\ref{fig8aa} respectively, for the
string melting version of the AMPT model. The general conclusions are similar, except the
magnitude of the $v_{2}$ and $v_{3}$ values are higher for string melting relative to
default case. Further one notices the difference in both  $v_{2}(p_{\rm T})$ and $v_{3}(p_{\rm T})$
for various collision configuration seems to have increased for string melting case compared to
default case. For the several configurations studied the  $v_{3}(p_{\rm T})$ in default AMPT model for U+U collisions 
are mostly below the corresponding values from Au+Au collisions, while for string melting case the U+U collision
$v_{3}(p_{\rm T})$ are mostly higher than the corresponding values from Au+Au collisions.

\section{SUMMARY}

In this study we have implemented the possibility of studying high energy collisions with deformed
Uranium nuclei within the framework of AMPT model. Experimental U+U collisions at 
around $\sqrt{s_{\rm {NN}}}$ = 200 GeV is planned for the year 2012 at the RHIC facility.
The Uranium nuclei is implemented by using a deformed Woods-Saxon profile and the projectile
and target Uranium nuclei are randomly rotated along the polar and azimuthal directions event-by-event 
with the probability distribution $\sin{\Theta}$ and uniform distribution for $\Theta$ and $\Phi$, respectively.
In the current work we have studied three specific configurations of U+U collisions for  $\sqrt{s_{\rm {NN}}}$ = 200 GeV,
based on the choice of the polar, azimuthal angles of the two nuclei and the impact parameter direction. The results from these
collisions have been compared to U+U collisions with no specific choice of orientation and Au+Au collisions
at the same beam energy.  For most of the observables studied we present the results for both default and 
string melting configurations of the AMPT model.

The charged particle multiplicity and charged particle transverse energy is found to be about 15\%--35\% higher 
for the different U+U configurations relative to Au+Au collisions. For string melting version the multiplicities
are higher by about 8\% for U+U collisions compared to default model, whereas for the transverse energy they are
lower by about 10\% compared to default. The average transverse momentum for charged particles increases with
number of participating nucleons for the default case, whereas they saturate for central collisions in case of
the string melting version. This is perhaps due to additional partonic interactions and the quark coalescence process 
in the string melting model relative
to the default case.  Within the different configurations studied, the multiplicity, 
transverse energy and average transverse momentum at midrapidity are largest for the tip-to-tip configuration.
This observation may be used in future to select events of particular configuration for U+U collisions.

The average eccentricity and fluctuations in eccentricity shows a rich dependence on collision centrality. 
The side-on-side configuration posses the maximum eccentricity and minimum eccentricity fluctuations
as a function of collision centrality among the configurations studied. On the contrary, the body-to-body configuration
has the minimum eccentricity and maximum eccentricity fluctuations for peripheral and mid-central collisions.
The tip-to-tip configuration has minimum eccentricity and maximum eccentricity fluctuations for central collisions.
We did not observe large variation in triangularity and its fluctuations for different configurations of U+U
collisions studied or between U+U collisions and Au+Au collisions. 

The variation of eccentricity and its fluctuation gets reflected in dependence of   $v_{2}$ and $v_{3}$  as
a function of collision centrality. In both the default and string melting versions of the model, the 
side-on-side configuration has maximum  $v_{2}$ compared to all other configurations and Au+Au collisions
studied for all collision centrality. The $v_{2}$ is smallest for the body-to-body configuration in peripheral
and mid-central collisions while it is minimum for tip-to-tip configuration in central collisions.
For peripheral collisions the $v_{\rm 2}$ in U+U  can be about 40\% larger than in Au+Au
whereas for central collisions it can be a factor 2 higher depending on the collision configuration.
These features in $v_{2}$ closely follow the features in the $\ecc$.  
The $v_{3}$  does not show much dependence on U+U configuration for 
AMPT calculations in default version, however in string melting version, it seems the body-to-body and
tip-to-tip configurations have higher $v_{3}$ values as a function of collision centrality compared to
other cases studied. The $p_{T}$ dependence of  $v_{2}$ is presented
for 80 $<$ $N_{\rm {part}}$ $<$ 120 and  160 $<$ $N_{\rm {part}}$ $<$ 200.
We observe a U+U side-on-side configuration has largest $v_{2}$ while U+U body-to-body configuration
has the smallest $v_{2}$. For the other configurations, tip-to-tip, U+U and Au+Au the values are similar. 
A more clearer  $p_{T}$ dependence of $v_{3}$ is observed in string melting case compared to default case. 
In the string melting case, the $v_{3}$ of tip-to-tip and body-to-body are similar and higher than 
U+U, U+U side-on-side and Au+Au collisions.

The future scope of our study includes studying the effect of jet-quenching and jet-medium interactions via 
dihadron correlations for different configurations of U+U collisions. Furthermore we plan to use the AMPT model
to study the most effective way to select various configuration in U+U collisions in an experiment.

\noindent{\bf Acknowledgments}\\
This work is supported by
the DAE-BRNS project grant No. 2010/21/15-BRNS/2026.

\end{document}